\documentclass[12pt,letter]{article}
\usepackage{longtable}
\usepackage{amsmath}
\usepackage{amsfonts}
\usepackage{multirow}
\usepackage{float}
\usepackage{lscape}
\usepackage[hang,bottom]{footmisc}
\usepackage{scalefnt}
\usepackage[margin=0pt,font=footnotesize,singlelinecheck=false]{caption}
\usepackage{amsthm}
\usepackage{comment}

\usepackage{booktabs}
\usepackage{tabularx}
\usepackage{colortbl}
\usepackage{bbm}
\usepackage{chapterbib}
\usepackage{natbib}
\usepackage{har2nat}
\usepackage[doublespacing]{setspace}
\usepackage{graphicx}
\usepackage{titlesec}
\titlelabel{\thetitle.\quad}
\titleformat{\subsection}
  {\normalfont\Large\itshape}{\thesubsection}{1em}{}

\definecolor{DarkBlue}{rgb}{0,0.08,0.45}
\usepackage{ifpdf}
\ifpdf

\usepackage[pdftex=true, colorlinks=true, citecolor=black, linkcolor=black]{hyperref}
\else

\usepackage[hyperfootnotes=true]{hyperref}
\fi

\usepackage{tabu}

\usepackage[top=2.5cm, bottom=2.5cm, left=2.5cm, right=2.5cm]{geometry}
\definecolor{orange}{rgb}{1,0.5,0}

\newtheorem{prop}{Proposition}
\newtheorem{assumption}{Assumption}
\newtheorem{coroll}{Corollary}
\usepackage{tikz}
\usetikzlibrary{bayesnet}

\usepackage{todonotes}

\usepackage[title]{appendix}

\definecolor{noteColor}{rgb}{0.75,0.1,0}

\begin{document}

\nocite{Andersen_Dobrev_Schaumburg_2014}
\nocite{Barndorff_Shephard_2002}
\nocite{Barndorff_Shephard_2005_accurate}
\nocite{Barndorff_Shephard_2005_variation}
\nocite{Barndorff_Shephard_Winkel_2006}
\nocite{Harvey_Ruiz_Shephard_1994}
\nocite{Shephard_1996}
\nocite{Liu_West_2001}

\begin{sloppypar}
\title{A Randomized Missing Data Approach to Robust \newline Filtering and Forecasting}

\date{September 30, 2022}

\author{Dobrislav Dobrev{$^\dagger$} \and Derek Hansen\thanks{\scriptsize University of Michigan,
    Department of Statistics, Ann Arbor, MI, 48109. Email: dereklh@umich.edu. This author acknowledges support from the National Science Foundation Graduate Research Fellowship Program under grant no. 1256260. Any opinions,
findings, and conclusions or recommendations expressed in this material are those of the
author(s) and do not necessarily reflect the views of the National Science Foundation.
 } \and Pawe{\l} J.
  Szersze\'{n}\thanks{\scriptsize Board of Governors of the Federal Reserve
    System, 20th St. and Constitution Ave. NW, Washington, DC 20551. Emails:
    dobrislav.p.dobrev@frb.gov and pawel.j.szerszen@frb.gov.
    This article represents the views of the authors, and should not be interpreted as reflecting the views of the Board of Governors of the Federal Reserve System or other members of its staff.}}

\maketitle

\begin{singlespace}
\begin{abstract}
\noindent {\footnotesize We put forward a simple new randomized missing data (RMD) approach to robust filtering of state-space models, motivated by the idea that the inclusion of only a small fraction of available highly precise measurements can still extract most of the attainable efficiency gains for filtering latent states, estimating model parameters, and producing out-of-sample forecasts. In our general RMD framework we develop two alternative implementations: endogenous (RMD-N) and exogenous (RMD-X) randomization of missing data. A degree of robustness to outliers and model misspecification is achieved by purposely randomizing over the utilized subset of data measurements in their original time series order, while treating the rest as if missing. The arising robustness-efficiency trade-off is controlled by varying the fraction of randomly utilized measurements. Our RMD framework thus relates to but is different from a wide range of machine learning methods trading off bias against variance. It also provides a time-series extension of bootstrap aggregation (bagging). As an empirical illustration, we show consistently attractive performance of RMD filtering and forecasting in popular state space models for extracting inflation trends known to be hindered by measurement outliers.}

\vspace*{.3in}



\end{abstract}
\end{singlespace}

\newpage
\section{Introduction}\label{Introduction}
State-space models play an important role in fields as diverse as engineering,
medicine, economics, and finance.
However, their empirical performance, particularly in forecasting applications, suffers when there is even a
small degree of model misspecification. This misspecification can be induced, for example, by
the presence of outliers in collected measurement data.
These problems have motivated
the development of methods for estimation of state-space models with
specific focus on robustness to complex data imperfections.
One popular approach imposes heavy-tailed distributional assumptions as in \citet{Durbin_Koopman_2000} or \citet{Harvey_Luati_2014}.
Another alternative adds thresholding or an outlier detection step as in \citet{Calvet_Czellar_Ronchetti_2015}, \citet{Crevits_Croux_2017}, \citet{Maiz_Molanes_Miguez_Djuric}, among others.

In this paper we propose a randomized missing data (RMD) framework for robust estimation of state-space models. Our framework achieves a degree of robustness to outliers and model misspecification by purposely randomizing the subset of included but possibly misspecifed or outlier contaminated data while treating the rest as if missing. This ensures that all available measurements would still get utilized but subject to a degree of downweighting. What makes our randomization approach work is that in many cases, most of the information for filtering latent states, estimating model parameters, and producing out-of-sample forecasts can be extracted from a small subset of available measurements. This is particularly true when the latent process is highly persistent through time. Thus, randomizing the inclusion of data points can achieve robustness to outliers and model misspecification without much efficiency loss. Moreover, this trade-off between robustness and efficiency loss can be controlled by tuning it to optimize out-of-sample forecasting metrics or, in general, by minimizing a loss function of interest.

A similar idea rationalizing the exclusion of data measurements was previously considered by \citet{Sims_2003} and later \citet{Sims_2011} who develop the notion of rational inattention. They show that, in the presence of explicit costs to collecting and processing measurements, it is optimal from a decision-theoretic point of view to not use all available data. We argue that the presence of measurement outliers and misspecification leads to a similar trade-off between bias and efficiency. Since the inclusion of all available observations can yield biased state and parameter estimates, the risk of including contaminated observations is an implicit cost to using more data. Hence, by inducing a degree of ``randomized inattention'' in our RMD framework, we can control for the trade-off between bias and efficiency subject to minimizing the loss function of interest.  As previously noted by \citet{Hamilton_1986}, among others, the total uncertainty of filtered states in state-space models can be decomposed into separate filter uncertainty and parameter uncertainty components. This decomposition further entails an inherent bias-efficiency trade-off with respect to the set of measurements taken into consideration in the presence of model misspecification, e.g. due to intermittent measurement distortions.

In Section \ref{sec:RMD_framework}, we lay out the general properties of the RMD framework in a Bayesian setting and then show two alternative implementations. The first one, denoted RMD-N, allows for endogenous randomization of missing data based on an indicator that is subject to a learning assumption. The second implementation, denoted RMD-X, is subject to a no-learning assumption with exogenous randomization of missing data. We establish that both RMD-N and RMD-X arise as special cases in our general RMD framework under specific choices of the generative model for missing observations. The imposed fraction of randomly utilized measurements plays the role of a regularization parameter controlling the arising robustness-efficiency trade-off in the presence of misspecified observations with favorable learning properties.

Our main theory result in Section \ref{sec::Bias_Variance_Decomposition} below is that the RMD framework introduces a new bias-variance trade-off controlled by the induced randomization rate. As a key distinction, the RMD framework limits data utilization to reduce prediction bias at the expense of larger variance, while a wide range of ML methods exploit bias-variance trade-off involving penalization over the entire data sample to reduce prediction variance at the expense of larger bias.
The RMD-X implementation of the RMD framework can further be viewed as a time-series extension of bootstrap aggregation (bagging), originally developed by \citet{Breiman_1996}. As a key distinction from bagging, RMD-X preserves time-series dependence by re-sampling without replacement while retaining the original time index of each observation by randomly drawing only induced missing values in each re-sampled path.

To illustrate the empirical performance of our approach, in Section \ref{sec:empirical} we consider state space models for extracting inflation trends and document favorable performance of our RMD framework and the resulting out-of-sample forecasts. We first apply RMD to the classical unobserved components (UC) model. It has been well documented by \citet{Stock_Watson_2007} and \citet{Stock_Watson_2016}, among others, that the forecasting performance of the UC model is hindered by the presence of time-variation in the precision of inflation rate measurements and additional measurement distortions due to outliers. We therefore study an RMD-N and RMD-X augmented versions of the classical UC model vis-\`{a}-vis the unobserved components/stochastic volatility outlier-adjustment (UCSVO) model that was put forward by \citet{Stock_Watson_2016} as another way to minimize the detrimental impact of outliers by way of subjecting them to particular distributional assumptions.

We find that our RMD-N and RMD-X augmented UC model that avoids overfitting via ``randomized inattention'' to the available measurements can offer modest improvements over the UCSVO model in terms of out-of-sample forecasting performance, especially at longer 8- and 12- month horizons. We conclude that the favorable forecasting results obtained when applying RMD filtering to the UC model underscore the findings in \citet{Stock_Watson_2007} and \citet{Stock_Watson_2016} regarding the need for time-series inflation forecasts to address the presence of potentially complex inflation measurement imperfections. We also show improvement in mean squared forecast error (MSFE) over a UC model with t-distributed innovations without stochastic volatility (UC-T), demonstrating that the filtering and forecasting benefits offered by RMD cannot be attained by simply imposing a heavier-tailed measurement distribution.

We further extend the empirical study of RMD extraction of inflation trends by considering also a more general AR model for the unobserved inflation component which allows for reversion to a long-run mean either estimated from the data or fixed at the target rate of $2\%$ as a way to more directly reflect central bank inflation targeting that has been in place for many years. The much superior forecasting performance we document when the mean is fixed to the \textit{a priori} $2\%$ inflation target in place suggests that inflation targeting has been effective. The application of the RMD framework can be beneficial also after imposing additional economic information about the underlying model. Both when applied to the classical UC and better-performing AR models with fixed inflation target mean our RMD approach to robust filtering and forecasting offers attractive empirical performance gains and ease of implementation in comparison to existing alternatives.

\section{The RMD Modelling Framework} \label{sec:RMD_framework}

For $t \ge 1$, the following equations characterize the Randomized Missing Data (RMD) modelling framework:
\begin{align}
  x_0                  &\sim \mu_\theta(\cdot)                        \label{eq::SSM_stationary} \\
  x_t | x_{t-1},\theta  &\sim g_\theta(\cdot | x_{t-1},\theta              \label{eq::SSM_state}) \\
  y_t |x_t,\theta &\sim f_\theta(\cdot |x_t)              \label{eq::SSM_obs_prec} \\
  {y}_t^{(j)}      &\sim f_{j} (\cdot | \hat y^{t-1})       \label{eq::SSM_junk_value} \\
  C_t                   &\sim  P(. | \beta)                 \label{eq::SSM_junk_indicator} \\
  \hat {y}_t           &= \begin{cases} \label{eq::SSM_missing_cond}
    y_t ~ \text{if} ~ C_t = 1 \\
    y_t^{(j)} ~ \text{if} ~ C_t = 0
  \end{cases}
\end{align}

\begin{figure}[t]
  \begin{center}
%
%
%


\begin{tikzpicture}





  \matrix[row sep=5mm, column sep=1cm] (LDA)
  { %
  \node[latent] (xtm1) {$x_{t-1}$}; & \node[factor][label=above:$g_\theta$] (theta1) {$ $}; & \node[latent] (xt) {$x_t$}; & \node[factor][label=above:$g_\theta$] (theta2) {$ $}; & \node[latent] (xtp1) {$x_{t+1}$}; \\
  & & \node[factor][label=right:$f_\theta$] (thetaf) {$ $}; & & \\
                                    & & \node[latent]  (yt2)      {$y_t$} ; & & \\
  & \node[latent] (ytj) {$y_{t}^{(j)}$};  &\node[factor][label=right:$C_t \sim P(. | \beta)$] (ct) {$ $}; \\
  \node[obs] (ythatp) {$\hat y_{t-1}$} ; & & \node[obs] (ythat) {$\hat y_{t}$} ; & &  \\
  };

  \factoredge{xtm1} {theta1} {xt};
  \factoredge {xt} {theta2} {xtp1};
  \factoredge {xt} {thetaf} {yt2};
  \factoredge {yt2} {ct} {ythat};
  \factoredge {ytj} {ct} {ythat};
  \edge {ythatp} {ytj};
  %
  %

\end{tikzpicture}

  \end{center}
  \caption{State space model with observed measurement $\hat{y}_{t}$ that, conditional on the value of the auxiliary indicator $C_t$, can be possibly corrupted/uninformative and given by $y_t^{(j)}$ or a correctly-specified one and given by $y_{t}$. Shaded nodes are the observed data. \label{fig:diagram_SSM}}
\end{figure}
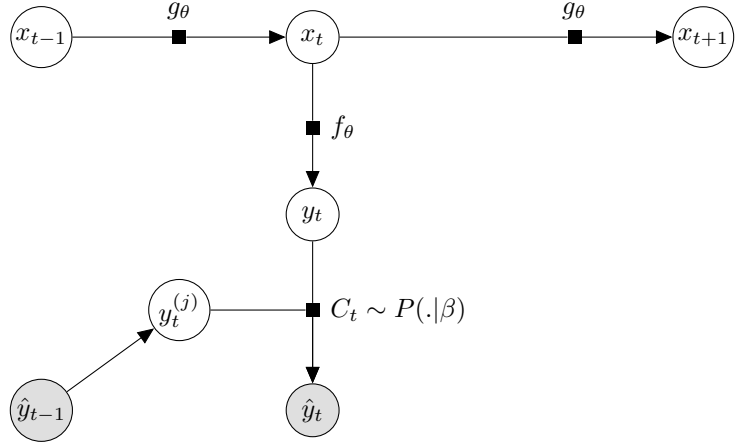

As in a standard state-space model, $x_t \in \mathbb R^{d_x}$ is a latent Markov process parameterized by $\theta \in \Theta$ with an initial distribution $\mu_\theta$ (\autoref{eq::SSM_stationary}) and a transition kernel $g_\theta$ (\autoref{eq::SSM_state}).
The observation $y_t \in \mathbb R^{d_{y}}$ contains information about the latent states $x_t$ through the kernel $f_\theta$ (\autoref{eq::SSM_obs_prec}).
However, unlike a standard state-space model, we further suppose that $y_t$ is only observed a fraction of the time as specified by the three additional equations (\ref{eq::SSM_junk_value})-(\ref{eq::SSM_missing_cond}).
Specifically, we observe $\hat y_t$, which equals the correctly-specified $y_t$ only when the auxiliary indicator variable $C_t \in \{0, 1\}$ equals $1$ (\autoref{eq::SSM_missing_cond}).
In the case that $C_t = 0$, the observed $\hat y_t$ instead equals a corrupted value $y_j \in \mathbb R^{d_y}$ from the \textit{alternative kernel} $f_j$ which, conditional on previous observations $\hat y^{t-1}$, is independent of the model parameters $\theta$, the latent process $x_t$, and current and future realizations of $y_t$.
Unlike a typical missing data problem, the observed $\hat y_t$ has a recorded value at every index from $1$ to $T$. However, based on the structure of the model, if the true $C^T$ were known, then the framework would simply reduce to standard estimation with missing data, since the $y_t^{(j)}$ have no additional information about $\theta$ or $x_t$.
Figure \ref{fig:diagram_SSM} provides a block-diagram illustration of the model with observed measurement $\hat{y}_{t}$ that, conditional on the value of the auxiliary indicator $C_t$, can be possibly corrupted/uninformative and given by $y_t^{(j)}$ or a correctly-specified one and given by $y_{t}$.

The goal of adopting the RMD modelling framework is to optimize the trade-off between the gain of including correctly-specified observations and the loss of including misspecified observations.
This is performed through controlling the distribution of $C_t$ through the parameter $\beta$.
Although the distribution of $C_t$ can be general, the simplest case is to assume the $C_t$ are i.i.d. with distribution $Bernoulli(\beta), ~ \beta \in [0,1]$.
In this formulation, $\beta = 1$ corresponds to including all measurements as if correctly specified, while the limiting case of $\beta = 0$ has no inclusion of data at all.
For $\beta \in (0, 1)$, the level of $\beta$ reflects the degree to which the observed $\hat y^T$ are included via randomization inducing downweighting  of the observed $\hat y^T$ to robustify model estimates and forecasts. As such, the three additional equations (\ref{eq::SSM_junk_value})-(\ref{eq::SSM_missing_cond}) characterizing our RMD framework can be viewed as an add-on to any standard state space model specification (\ref{eq::SSM_stationary})-(\ref{eq::SSM_obs_prec}) for the purposes of robust filtering and forecasting in the presence of measurement outliers and model mis-specification even without knowledge of the exact form of such imperfections.

In addition to the prior distribution for $C^T$, the nature of the induced missing data randomization depends on the distribution of the alternative measurement $f_j$. In general, the alternative measurement at time $t$ is assumed to be independent of model parameters $\theta$ and states $x^T$ conditional on observations up to time $t-1$.
Using the general case, we are able to derive the posterior distribution $P(x^T, \theta, C^T | \hat y^T, \beta)$ sequentially, showing that at each $t$ the distribution is a mixture between including each $\hat y_t$ as if it were equal to its informative counterpart $y_t$ and treating the measurement as if it were missing. The model posterior distriubution is given by
\begin{equation} \label{eq::posterior_full_maintext}
  P(x^T, \theta | \hat y^T, \beta) = \prod_{t=1}^T \Big (\hat \beta(\hat y^t) \frac{f_\theta(\hat y_t | x_t)} {F(\hat y_t | \hat y^{t-1}, \beta)} + (1 - \hat \beta(\hat y^t) \Big ) g_\theta(x_t | x_{t-1}) \mu_\theta (x_0) P(\theta)
\end{equation}
Here, $\hat \beta(\hat y^t)$ is shorthand for the filtered posterior probability $P(C_t =1 | \hat y^t, \beta)$ derived in Appendix \ref{sec:RMD_posterior}, and ${F(\hat y_t | \hat y^{t-1}, \beta) = \int_{x_t, \theta} f_\theta (\hat y_t | x_t) dP(x_t, \theta | \hat y^{t-1}, \beta)}$ is the one-step-ahead predictive density of $y_t$ evaluated at $\hat y_t$. A detailed derivation of the posterior distribution can be found in Appendix \ref{sec:RMD_posterior}.

Given this posterior form in general, a natural question that arises is whether we can specify $f_j$ in a way that avoids additional parametrization and leaves $\beta$ as the only parameter controlling the arising robustness efficiency trade-off. We show two alternative assumptions which lead to two such forms for $f_j$ and respective alternative RMD implementations based on endogenous randomization (RMD-N) or exogenous randomization (RMD-X).
In the RMD-N specification, we assume that the missing indicator $C_t$ cannot be learned from past or present observations $\hat y_t$, but can be learned from future observations. From this assumption, we can derive the exact form that the alternative distribution $f_j$ must take.
This approach has favorable learning properties, because one can infer where in the past $E(C_t | \hat y^T)$ is low, and also imposes neglible computational cost compared to the original state-space model defined by equations (\ref{eq::SSM_stationary})-(\ref{eq::SSM_obs_prec}).

In contrast, in the RMD-X specification, we assume that the missing indicator $C_t$ cannot be inferred even after observing all data points.
We show that this leads to another exact form for $f_j$ which induces a posterior distribution which randomizes over $P(C^T|\beta)$ solely as a function of $\beta$.

\subsection{RMD-N: Endogenous Randomization} \label{subsec:RMD_N_Assumption}
The RMD-N approach models the indicator variable $C^T$ endogenously within a Bayesian hierarchical model using the relationships outlined in Equations (\ref{eq::SSM_stationary}) - (\ref{eq::SSM_missing_cond}).
This method requires an explicit specification of the alternative distribution $f_{j}$. As indicated above, there is much flexibility in how $f_j$ is chosen. In our RMD-N framework we assume that $y_t^{(j)}$ has distribution given by the one-step-ahead
predictive density $F(\cdot | \hat y^{t-1}, \beta)$ and is conditionally independent of all other past, present, and future realizations. This is equivalent to assuming the contaminated measurements should be indistinguishable from the true measurements at the time of observing them, but allowing for the possibility that future values may be informative about $C_t$. This particular choice of $f_j$ pairs well with existing state-of-the-art Sequential Monte Carlo (SMC) techniques for jointly estimating parameters and states such as SMC$^2$ from \citet{Chopin_Jacob_Papas_2013}. This is because at each point in time the latent $C_t$ can be marginalized out, and $f_j(\hat y_t)$ can be calculated as a density forecast over the available filtered particles  sampled from $P(x_{t-1}, \theta | \hat y^{t-1}, \beta)$.

We start by supposing that $C_t$ is not inferable with information up to time $t$.
Specifically, $P(C_t = 1| \hat y^t, \beta) = P(C_t = 1| \beta) = \beta$. Following Equation (\ref{eq::post_ct2}), this assumption implies that $f_j(\hat y_t | \hat y^{t-1}, \beta) = F(\hat y_t |\hat y^{t-1}, \beta)$, which is the one-step-ahead predictive density evaluated at the observed data point. The posterior distribution in Equation (\ref{eq::posterior_sequential}) can also be further simplified
\begin{equation} \label{eq:RMDN_fj_is_f_posterior}
  P(x^t, \theta | \hat y^t) = (\beta \frac{f(\hat y_t | x_t, \theta)}{F(\hat y_t | \hat y^{t-1}, \beta)} + (1 - \beta)) g_\theta(x_t | x_{t-1})P(x^{t-1}, \theta | \hat y^{t-1})
\end{equation}
Because the contaminated measurement $y_t^{(j)}$ is indistinguishable in distribution from the true measurement $y_t$  with present information, the inclusion of the observed $\hat y_t$ at the time of observation is only weighted by the prespecified prior probability $\beta$.
However, while the filtered probability $P(C_t = 1 | \hat y^t, \beta)$ equals $\beta$, the smoothed probability $P(C_t = 1 | \hat y^T, \beta)$ will not necessarily equal $\beta$.
This is because future realizations of $\hat y_t$ will be informative about $C_t$.
Thus, at the end of estimation, the inclusion and weighting of a measurement at a particular point in time will depend on the extent to which it is corroborated by future observations.
These smoothed probabilities can also be of interest to the researcher in diagnosing where the original model breaks down.

\subsection{RMD-X: Exogenous Randomization} \label{subsection:RMD_X}

The data randomization prescribed by RMD-X fixes the weight on each subset of the data beforehand.
More specifically, the RMD-X approach exogenously draws each path $C^T$ from some
prespecified distribution satisfying the condition $P(C_t = 1) = \beta$. Parameter estimates and forecasts are
calculated for each $C^T$ as they would be in the usual missing-data context.
The estimates and forecasts are then averaged across paths, making them more robust to data imperfections at particular points.
Because the missing data randomization is performed outside of the estimation process, there is no learning about the indicator $C_t$.

We derive a specific choice of $f_j$ in our general RMD framework which elicits a posterior distribution equivalent to the purely exogenous randomization of RMD-X. Instead of assuming that only $C_t$ cannot be inferred with information up to time $t$, as in the case of RMD-N, we now assume that the entire path $C^t$ up to time $t$ cannot be learned from the data.
This leads to another explicit formulation of $f_j$ inducing a mixture distribution over all posterior distributions conditional on $C^T$ which are weighted only by the prior weight assigned to $C^T$.
Thus, we are able to justify averaging estimates over repeat data copies with randomly induced missing values as a key simplifying feature of the RMD-X approach.

To obtain an RMD-X implementation that fixes the weight on each subset of the data beforehand we set $f_j$ as follows:
$$
f_j(\hat y_{\{t: C_t = 0\}} | C^T, \hat y_{\{t: C_t = 1\}}) = P_y(\hat y_{\{C_t=0\}} | \hat y_{\{C_t=1\}})
$$

where $P_y$ is the uncontaminated observation distribution.
With this selection of $f_j$, $\hat y^T$ is independent of $C^T$:
\begin{align*}
    P(\hat y^T | C^T) = P_y(\hat y_{\{C_t = 1\}}) P_y(\hat y_{\{C_t=0\}} | \hat y_{\{C_t=1\}})  = P_y(\hat y^T)
\end{align*}
From this, $P(C^T | \hat y^T, \beta) = P(C^T | \beta)$, and we obtain the posterior as follows:
\begin{align*}
    P(x^T, \theta | \hat y^T, \beta) &= \sum_{C^T} P(C^T | \hat y^T, \beta) P(x^T, \theta | C^T, \hat y^T) \\
    &= \sum_{C^T} P(C^T | \beta) \frac{P(x^T, \theta) P(\hat y^T | C^T, x^T, \theta)}{P(\hat y^T | C^T)} \\
    &= \sum_{C^T} P(C^T | \beta) \frac{P(x^T, \theta) \prod_{t : C_t=1} f(\hat y_t |  x_t, \theta) f_j(\hat y_{\{C_t=0\}} | \hat y_{\{C_t=1\}})} {P(\hat y_{\{C_t=1\}}) f_j(\hat y_{\{C_t=0\}} | \hat y_{C_t=1})} \\
    &= \sum_{C^T} P(C^T | \beta) \frac{P(x^T, \theta) \prod_{t : C_t=1} f(\hat y_t |  x_t, \theta)} {P(\hat y_{\{C_t=1\}})} \\
    &= \sum_{C^T} P(C^T| \beta) P(x^T, \theta | \hat y_{\{C_t = 1\}})
\end{align*}

Thus, if the estimator of interest is the posterior mode, $f_j$ integrates out and this leads to RMD-X implementation equivalent to a weighted mixture over all possible  $C^T$:
\begin{equation} \label{eq::posterior_pure_random}
  \begin{split}
P(x^T, \theta | \hat y^T, \beta) = \sum_{C^T} P(C^T| \beta) P(x^T, \theta | \hat y_{\{C_t = 1\}})
\end{split}
\end{equation}

This result justifies casting RMD-X in simpler terms allowing for MLE estimation with the use of bootstrap-like aggregation over data replicas that preserve time series dependence by randomly drawing missing values and retaining the original time series order of the rest.
More formally, for a designated weighting scheme $P(C^T | \beta)$, the two-step algorithm for computing the RMD-X filtered states and parameter estimates is as follows:

\noindent \underline{Step 1}: Path-by-path filtering and forecasting for each path $C_i^T \in \{0, 1\}^T$:

  \begin{itemize}
    \item[$-$] Assume $\hat y_{\{t : C_{t, i} = 1\}} \overset D = y_{\{t : C_{t, i} = 1\}}$ conditional on $\theta, x^T$, with $\hat y_{\{t : C_{t, i} = 0\}}$ missing at random.
    \item[$-$] Obtain parameter estimates $\tilde \theta_i := \tilde \theta (\hat y_{\{t : C_{t, i} = 1\}})$.
    \item[$-$] Obtain filtered and forecasted states $\tilde x^{T+h}_{i} = \tilde x^{T+h} (\hat y_{\{t : ~ C_{t, i} = 1\}} , \tilde \theta_i)$.
  \end{itemize}

\vspace{0.05in}

\noindent \underline{Step 2}: Aggregation across all paths $C_i^T \in \{0, 1\}^T$:
\begin{itemize}
  \item[$-$] Obtain RMD-X parameter estimates $\bar \theta = \bar \theta (\hat y^T, \beta) = \sum_i \tilde \theta_i P(C_i^T | \beta)$.
  \item[$-$] Obtain RMD-X filtered and forecasted states $\bar x^{T+h} = \bar x^{T+h} (\hat y^T, \beta) = \sum_i \tilde x^{T+h}_{i} P(C_i^T | \beta)$.
\end{itemize}

The estimates calculated in RMD-X can be in principle any function of the data, but generally $\tilde \theta_i$ will either be the maximum-likelihood estimator of $\theta$ or a statistic of the posterior distribution if performing a Bayesian analysis. Likewise, if the goal is to minimize squared-error loss, then the estimated $\tilde x_{t+h, i}$ and forecasted $y_{t+h, i}$ will generally be calculated as expectations over the given subset of data and the estimated $\tilde \theta_i$.

The weighting scheme chosen for each path $C^T$ is flexible. The most basic scheme one could use is sampling each $C_t \sim Bernoulli(\beta)$ independently. However, especially for low $\beta$, this design assigns significant probability to using a low number of observations, which can lead to identifiability issues due to an insufficient number of observations. Therefore, a more practical approach is to fix the number of observations at $[\beta T]$, where $[x]$ is the nearest integer to $x$. This leads to the following distribution for $C^T$:

\begin{equation} \label{eq:RMDX_weighting}
  P(C^T) = \begin{cases} {T \choose [\beta T]}^{-1} ~ \text{if} ~ |C^T| = [\beta T] \\ 0 ~ \text{otherwise} \end{cases}
\end{equation}

In practice, since the sample space of $C^T$ is growing exponentially, it is infeasible to calculate the expectation across all possible $C^T$. Therefore, in applications, the expectation can be replaced with an average over Monte Carlo samples from $P(. | \beta)$. While this introduces some stochastic noise to the estimation, in practice we find that this noise is small provided a large enough number of samples drawn.

\subsection{RMD-X versus RMD-N: Advantages and Disadvantages}

A key advantage to the RMD-X method is its simplicity by purposefully inducing and randomizing over missing data in the original model.
Focusing specifically on linear Gaussian state-space models, many software packages already have the built-in capability to handle missing data.
All that is left to the user is to implement the sampling scheme for $C^T$, which can be as simple as sampling from $\{1, \dots, T\}$ without replacement.
Finally, RMD-X also allows for model estimation, filtering and forecasting robust to data contamination without the need
to specify a specific model for that contamination.

However, because the missing data randomization is handled exogenously outside of the statistical model, one cannot learn about $C^T$ using RMD-X.
For forming robust predictions about $\hat y_t$, this is less important, but there may be interest in inferring which values of $E(C_t | \hat y^T)$ are particularly low.
Another realted drawback to both RMD-X and $f_j$ taken in the ``no-learning'' approach is that calculation of the likelihood requires conditioning on an entire path of $C^T$.

By contrast, in the case of RMD-N and $f_j$ with ``smoothed'' learning, $C_t$ can be marginalized out.
This means that $f_j(\cdot)$  can be estimated over the existing filtered particles of $x_{t-1}$ and $\theta $ already available in an SMC$^2$ estimation.
Hence, this choice of $f_j$ within RMD-N adds negligible cost to the run-time of the algorithm.

As such, RMD-X introduces significant computational overhead because it estimates the model independently for each draw of $C^T_i$. Hence, it becomes impractical with a large number of observations $T$. In cases where the original model has a tractable likelihood function, such as via the Kalman filter, this overhead is manageable.
However, many processes of interest follow a non-linear or non-Gaussian specification.
These more general models require stochastic estimation methods such as Sequential Monte Carlo (SMC) which are computationally expensive, making this additional overhead impractical. Further, RMD-X becomes unstable with small values of $\beta$ and a small number of observations $T$, where the sample size effectively used for estimation becomes too small for a model to be estimated.

Summing up, RMD-X offers a simple way to implement our RMD framework but cannot be applied in all scenarios, while RMD-N offers a tractable way to estimate models with very large $T$ or small values of $\beta$ as well as to learn about $C^T$ and improve conditionally the amount of utilized information from each available measurement.

\subsection{Bias-Variance Decomposition of Expected Prediction Error}\label{sec::Bias_Variance_Decomposition}

Both the RMD-N and RMD-X implementations of the RMD framework robustify the forecasts $\bar x_{T+h}(\beta)$ of future states $x_{T+h}$ by inducing randomization rate $\beta \in (0,1]$ between the model-implied kernel $f(.)$ and the alternative kernel $f_j(.)$ for the available measurements $y_t$ up to time $T$. Both implementations share two key features. First, as detailed in sections \ref{subsec:RMD_N_Assumption} and \ref{subsection:RMD_X} the choice of $f_j(.)$ is model-free in the sense of avoiding the need for introducing any other parameters except for the randomization rate $\beta$. Second, as we demonstrate in this section, under mild regularity conditions the bias in filtered and forecasted states can be reduced by shrinking the randomization rate $\beta$.

This leads to our main theory result below that the RMD framework introduces a new bias-variance trade-off controlled by the randomization rate $\beta$. As a key distinction, the RMD framework limits data utilization to reduce prediction bias at the expense of larger variance, while a wide range of ML methods exploit bias-variance trade-off involving penalization over the entire data sample to reduce prediction variance at the expense of larger bias.

Our theory builds on the following generic assumption on the prevalence and nature of outliers contaminating the available measurements as the sample size $T$ gets large.

\begin{assumption}\label{assumption::outlier_fraction} A fraction $m \in [0,1)$ of the observations are i.i.d. outliers, that need \textit{not} follow either the SSM observation equation (\ref{eq::SSM_obs_prec}) or the auxiliary RMD equation (\ref{eq::SSM_junk_value}), with $M \subset \{1, 2, ..., T\}$ denoting the subset of time indices of all observations contaminated by outliers and $M' = \{1, 2, ..., T\} \setminus M$ denoting the subset of time indices of all other uncontaminated observations.
\end{assumption}

It is instructive to establish bias reduction separately for RMD-X and RMD-N, subject to additional mild regularity assumptions.

\subsubsection{Bias Reduction with RMD-X}\label{RMDX_bias_reduction}

We consider the bias of RMD-X filtered and forecasted states  $\mathbb{E} \bar x^{T+h}(\hat y^T, \beta) - \mathbb{E} x^{T+h}$ using the notation from the two-step algorithm in section \ref{subsection:RMD_X} with indicator paths $C^T$ following the distribution from equation (\ref{eq:RMDX_weighting}).

\begin{prop}\label{Bias_RMDX_Beta}
Given Assumption \ref{assumption::outlier_fraction}, let $R = m \cdot T = |M|$ be the number of outliers among the observations $y^T$ for time indices $t \in M$ and $K = \beta T= |C_i^T| $ be the number of retained observations by each RMD-X draw of the indicator path $C_i^T$, $i=1,2,...,\binom{T}{K}$ with corresponding filtered and forecasted states $\tilde x^{T+h}_{i}$. Further assume that $R = o(T)$ and the biases associated with each indicator path $C_i^T$, $i=1,2,...,\binom{T}{K}$ are uniformly bounded: $\left| \mathbb{E} \tilde x^{T+h}_{i} - \mathbb{E} x^{T+h} \right|<B$ as $T$ gets large. Then the RMD-X filtered and forecasted states $\bar x^{T+h} (\hat y^T, \beta) = \frac{1}{\binom{T}{K}}\sum_i \tilde x^{T+h}_{i} $ would be asymptotically unbiased as $\beta = \frac{K}{T}$ becomes small when $T$ gets large:
\begin{equation}
\mathbb{E} \bar x^{T+h}(\hat y^T, \beta) - \mathbb{E} x^{T+h} \longrightarrow 0 ~~ \text{as} ~ \beta=\frac{K}{T} \rightarrow 0 ~ \text{with} ~ T \rightarrow \infty
\end{equation}

\end{prop}
\begin{proof}
The indices of $i=1,2,...,\binom{T}{K}$ of all indicator paths $\{ C_i^T \in \{0, 1\}^T ~ : ~ |\{t : C_{t, i} = 1\}| = K\}$ can be partitioned into two disjoint subsets $I_0 = \{ i\,:\, C_{t,i}=0 ~ \text{for all} ~ t \in M \}$ and $I_1 = \{ i\,:\, C_{t,i}=1 ~ \text{for some} ~ t \in M \}$ based respectively on whether the retained $K$ observations by $C_i^T$  contain outliers or not. Observe that $|I_0|=\binom{T-R}{K}$ and $|I_1|=\binom{T}{K}-\binom{T-R}{K}$. This leads to the following expression for the bias of the RMD-X filtered and forecasted states:

\vspace{-0.2in}
\begin{align*}
\mathbb{E} \bar x^{T+h} (\hat y^T, \beta) - \mathbb{E} x^{T+h} &= \frac{1}{\binom{T}{K}} \sum_i \mathbb{E} \tilde x^{T+h}_{i} - \mathbb{E} x^{T+h} \\
&= \frac{1}{\binom{T}{K}} \sum_{i \in I_0} \mathbb{E} \tilde x^{T+h}_{i}  +  \frac{1}{\binom{T}{K}} \sum_{i \in I_1} \mathbb{E} \tilde x^{T+h}_{i} - \mathbb{E} x^{T+h}\\
&= \biggl( \underbrace{\frac{1}{\binom{T}{K}} \sum_{i \in I_0} \mathbb{E} \tilde x^{T+h}_{i}}_{\text{No bias: $\binom{T-R}{K}$ terms}} - ~~ \frac{\binom{T-R}{K}}{\binom{T}{K}}\mathbb{E} x^{T+h} ~\biggr)  \\
&+  \biggl( \underbrace{\frac{1}{\binom{T}{K}} \sum_{i \in I_1} \mathbb{E} \tilde x^{T+h}_{i}}_{\text{Bias: $\binom{T}{K} - \binom{T-R}{K}$ terms}} - ~~ \frac{\binom{T}{K}-\binom{T-R}{K}}{\binom{T}{K}}\mathbb{E} x^{T+h} ~\biggr)
\end{align*}

The expression in the first bracket containing the no-bias terms would thus equal zero by construction. From this it follows:

\begin{align*}
\left| \mathbb{E} \bar x^{T+h} (\hat y^T, \beta) - \mathbb{E} x^{T+h} \right| &= \biggl| \frac{1}{\binom{T}{K}} \sum_{i \in I_1} \mathbb{E} \tilde x^{T+h}_{i} - ~~ \frac{\binom{T}{K}-\binom{T-R}{K}}{\binom{T}{K}}\mathbb{E} x^{T+h} ~\biggr| \\
&= \frac{1}{\binom{T}{K}} ~ \biggl| \underbrace{\sum_{i \in I_1} \left( \mathbb{E} \tilde x^{T+h}_{i} - \mathbb{E} x^{T+h} \right)}_{\text{$\binom{T}{K} - \binom{T-R}{K}$ terms}} ~\biggr|\\
&\leq \frac{\binom{T}{K}-\binom{T-R}{K}}{\binom{T}{K}} \cdot B \\
&= \left( 1~-~\frac{\binom{T-R}{K}}{\binom{T}{K}} \right) \cdot B \longrightarrow 0 ~~ \text{as} ~ \beta=\frac{K}{T} \rightarrow 0 ~ \text{with} ~ T \rightarrow \infty ~,~
\end{align*}

where the convergence to zero follows from

\begin{align*}
\frac{\binom{T-R}{K}}{\binom{T}{K}} &= \frac{(T-R)(T-R-1)...(T-R-K+1)}{T(T-1)...(T-K+1)} \longrightarrow 1 ~~\text{as} ~ R=o(T) ~ \text{with} ~ T \rightarrow \infty \\
\end{align*}

\end{proof}

\noindent {\bf Remark.} Similar reasoning implies that a degree of bias reduction can be attained also when $R=O(T)$ without anymore being able to guarantee that the bias would vanish asymptotically even if it can still be reduced.

\subsubsection{Bias Reduction with RMD-N}\label{RMDN_bias_reduction}

In addition to Assumption \ref{assumption::outlier_fraction}, we assume consistency of the RMD-N augmented model when the set $M$ of outliers is \textit{known} and inference is carried out based on the rest of correctly specified observations in the set $M'$. In addition, we also assume non-negligible bias when $M$ is unknown in the presence of outliers, i.e., that the outliers matter also asymptotically.

\begin{assumption}\label{assumption::consistency_and_bias} The choice of prior distributions $P(\theta)$ and $\mu_\theta (x_0)$ ensures consistent estimation of states and parameters in the model augmented with RMD given by equations (\ref{eq::SSM_stationary}) - (\ref{eq::SSM_missing_cond}), when the set $M$ of outliers is known and the outlier observations are treated as missing, i.e.,
\begin{equation} \label{eq::posterior_outliers_removed}
  \hat{P}(x^T, \theta | \hat y^T, \beta, M) = \prod_{t \in M'} \Big (\hat \beta(\hat y^t) \frac{f_\theta(\hat y_t | x_t)} {F(\hat y_t | \hat y^{t-1}, \beta)} + (1 - \hat \beta(\hat y^t)) \Big ) \prod_{t=1}^T g_\theta(x_t | x_{t-1}) \mu_\theta (x_0) P(\theta)
\end{equation}

In the opposite case when the set $M$ of outliers is unknown and the outlier observations are not removed, we further assume that for all $\beta \in (0,1]$ the estimated states and parameters are asymptotically biased in the model augmented with RMD-N whose posterior distribution is given by equation (\ref{eq::posterior_full}).
\end{assumption}

We can now derive our main result assessing the posterior bias of the RMD-N augmented model.

\begin{prop}\label{Bias_RMDN_Beta} Under Assumptions \ref{assumption::outlier_fraction} and \ref{assumption::consistency_and_bias}, for any given choice of $\beta \in (0,1]$ there exists a sample size $T'(\beta)$ that depends on the value of $\beta$ such that for all $T>T'$ the bias of h-step ahead predictions in the RMD-N augmented state-space model given by equations (\ref{eq::SSM_stationary}) - (\ref{eq::SSM_missing_cond}) is a locally increasing function of the mixing coefficient $\beta$, i.e., there exists $\delta$ such that the bias is increasing in the interval $(\beta-\delta,\beta+\delta)$.
\end{prop}
\begin{proof}
The posterior belief of $C_t=1$ given in equation (\ref{eq::post_ct1}), $\hat \beta(\hat y^t)$, is a strictly increasing function of $\beta$, which follows from a basic calculation of its first derivative w.r.t $\beta$ and its strictly positive value for $\beta \in (0,1]$. The state space model in equations (\ref{eq::SSM_stationary}) - (\ref{eq::SSM_missing_cond}) augmented by the RMD framework admits the following h-step ahead forecast density:
\begin{equation}
    P(x_{T+h}, x^T, \theta|\hat y^T, \beta) = \prod_{k=1}^h g_\theta(x_{T+k} | x_{T+k-1}) P(x^T, \theta | \hat y^T, \beta) ~,
\end{equation}
which follows directly from the definition of the state-space model. Hence, it is clear that to assess the bias of $x_{T+h}$ it suffices to assess the bias of the smoothed distribution of $x^T$.

The posterior in equation (\ref{eq::posterior_full}) can be rewritten, denoting outlier observations at time $t$ as $y_t^{(O)}$:
\begin{align*} 
  P(x^T, \theta | \hat y^T, \beta) = \prod_{t \in M} \Big( \hat \beta(y_t) \frac{f_\theta(y_t^{(O)} | x_t)} {F(y_t^{(O)} | \hat y^{t-1}, \beta)} + (1 - \hat \beta(\hat y^t)) \Big ) \cdot \\ \prod_{t \in M'} \Big (\hat \beta(y^t) \frac{f_\theta(y^t | x_t)} {F(y^t | \hat y^{t-1}, \beta)} + (1 - \hat \beta(\hat y^t)) \Big ) \prod_{t=1}^T g_\theta(x_t | x_{t-1}) \mu_\theta (x_0) P(\theta)
\end{align*}
Since the  $P(x^T, \theta | \hat y^T, \beta)$ is the only part of the integrand that depends on $\beta$ when calculating bias, and since it is continuously differentiable with respect to $\beta$, the bias is continuously differentiable with respect to $\beta$ following Leibnitz's rule. Using Assumptions \ref{assumption::outlier_fraction} and \ref{assumption::consistency_and_bias} for the two last terms there exists $T'(\beta)$ such that for all $T>T'$ the bias contribution of these terms can be made arbitrarily small, and thus its derivative w.r.t. to $\beta$ can be made arbitrarily small (in absolute value, as the bias is decreasing with $\beta$ for these terms). However, the bias increases with $\beta$ for the first term and the bias contribution of the first term does not vanish with a large sample size by Assumption \ref{assumption::consistency_and_bias}, i.e., even for large $T$ the bias is sensitive to the presence of outliers. Hence, for a given $\beta$ and sufficiently high $T>T'(\beta)$ the posterior distribution's bias is locally increasing in $\beta$.
\end{proof}

\begin{coroll}
Under Assumptions \ref{assumption::outlier_fraction} and \ref{assumption::consistency_and_bias} and for the choice of $T'$ from Proposition \ref{Bias_RMDN_Beta}, the bias of RMD-N filtered and forecasted states can be locally decreased by reducing the value of $\beta$.
\end{coroll}
\begin{proof}
The result follows directly from Proposition \ref{Bias_RMDN_Beta}.
\end{proof}

\subsubsection{Bias-Variance Decomposition with RMD}\label{subsec::Bias_Variance_Tradeoff}

The obtained results in Propositions \ref{Bias_RMDX_Beta} and \ref{Bias_RMDN_Beta} establish that the RMD framework offers a way to reduce prediction biases by lowering the randomization rate ($\beta<1$). On the flip side, akin to sub-sampling, restricting data utilization always comes at the expense of larger variance relative to utilizing the full sample ($\beta=1$). This leads to our main theory result that the RMD framework introduces a new bias-variance trade-off controlled by the randomization rate $\beta$ as detailed by Proposition \ref{eq::MSE_decomposition} below.

\begin{prop}{Bias-Variance Decomposition}\label{eq::MSE_decomposition}
			\begin{align*}
			\mathbb{E} \Bigl(\bar x_{T+h}(\beta,\bar\theta(T,\beta)) - x_{T+h} \Bigr)^2 \nonumber &  = \underbrace{\mathbb{E} \Bigl(\bar x_{T+h} (\beta,\bar\theta(T,\beta)) - \mathbb{E}  \bar x_{T+h}(\beta,\bar\theta(T,\beta)) \Bigr)^2 }_{\text{Reducible variance term} ~ \downarrow ~ \text{as} ~ \beta ~  \uparrow}  + \\
			& + \underbrace{\,\Bigl( \mathbb{E}  \bar x_{T+h}(\beta,\bar\theta(T,\beta)) \,  - \mathbb{E}  x_{T+h}  \Bigr)^2 \,}_{\text{Reducible bias term} ~ \downarrow ~ \text{as} ~ \beta ~  \downarrow}  + \\
			& + \underbrace{\mathbb{E} \Bigl( \mathbb{E} x_{T+h} \,  - x_{T+h}  \Bigr)^2 }_{\text{Irreducible variance term}} \,
			\end{align*}
where $\bar\theta(T,\beta)$ denotes either the MLE parameter estimate for RMD-X, or is a random variate from the RMD-N model's posterior distribution $P(\theta|Y^T)$ and all expectations reflect the distribution of measurements and latent states given information up to time $T$.
\end{prop}

\begin{proof}
The result follows directly from the following standard decomposition:
			\begin{align*}
			\mathbb{E} \Bigl(\bar x_{T+h}(\beta,\bar\theta(T,\beta)) - x_{T+h} \Bigr)^2 & = \mathbb{E} \Bigl(\,\bigl(\bar x_{T+h} (\beta,\bar\theta(T,\beta)) - \mathbb{E}  \bar x_{T+h}(\beta,\bar\theta(T,\beta)) \bigr) \,  + \\
			& +\, \bigl(\mathbb{E}  \bar x_{T+h}(\beta,\bar\theta(T,\beta)) \,  - \mathbb{E}  x_{T+h}\bigr) \, + \, \bigl(\mathbb{E}  x_{T+h} \,  - x_{T+h}\bigr)  \, \Bigr)^2 \nonumber \\
			& = \mathbb{E} \Bigl(\bar x_{T+h} (\beta,\bar\theta(T,\beta)) - \mathbb{E}  \bar x_{T+h}(\beta,\bar\theta(T,\beta)) \Bigr)^2  + \\
			&+ \,\Bigl( \mathbb{E}  \bar x_{T+h}(\beta,\bar\theta(T,\beta)) \,  - \mathbb{E}  x_{T+h}  \Bigr)^2 \, + \\
			& + \mathbb{E} \Bigl( \mathbb{E} x_{T+h} \,  - x_{T+h}  \Bigr)^2  \,
			\end{align*}
The first term reflects the reducible variance of RMD predictions by increasing $\beta$ towards its upper limit of $1$ corresponding to full-sample inference. The second term reflects the reducible bias of RMD predictions by decreasing $\beta$ as established by Propositions \ref{Bias_RMDX_Beta} and \ref{Bias_RMDN_Beta}. The last term reflects the DGP-implied irreducible variance of latent state predictions.
\end{proof}

\vspace{0.2in}
\noindent {\bf Remark.} The obtained bias-variance decomposition has the following implications for the use of RMD filtering and forecasting depending on whether outliers are present or not:
\begin{enumerate}
    \item Correctly specified model with large sample size $T$ in the absence of outliers \\
    In the correctly specified case parameter posterior consistency of the state space model when using the full data sample ($\beta=1$) is implied by the MLE parameter consistency under mild additional assumptions along the lines of \citet{Douc_2020}. Hence, for sufficiently large $T$ the marginal benefit of further reducing the bias term when decreasing $\beta$ as shown in Propositions \ref{Bias_RMDX_Beta} and \ref{Bias_RMDN_Beta} is smaller than the marginal cost of potentially increasing the variance term due to missing observations. This, in turn, implies that under the true model, and for sufficiently large $T$, the optimal value of $\beta$ is $1$. As such, RMD filtering and forecasting would not yield any improvement.
    \item Mis-specified model with large sample size $T$ in the presence of outliers \\
    In the presence of outliers, provided that model misspecification  induces non-negligible bias when utilizing the full sample ($\beta=1$), for sufficiently large $T$ the marginal benefit from decreasing $\beta$ as shown in Propositions \ref{Bias_RMDX_Beta} and \ref{Bias_RMDN_Beta} can be larger than the marginal cost of potentially increasing the variance term due to missing observations. This, in turn, implies that under a mis-specified model, and for sufficiently large $T$, the optimal $\beta$ is strictly less than $1$. As such, RMD filtering and forecasting would yield an improvement.
\end{enumerate}

In the machine learning literature such bias-variance decomposition is known to arise in many prediction problems under square loss function involving penalization over the entire data sample with the goal to reduce prediction variance at the expense of larger bias. By contrast, RMD filtering and forecasting is able to reduce prediction bias at the expense of larger variance by limiting data utilization. It should therefore be possible to apply RMD filtering and forecasting also in combination with penalty-based ML methods. As a key feature of RMD, the variance of $\bar x_{T+h}(\beta,\bar\theta(T,\beta))$ decreases as the randomization rate $\beta$ goes up allowing the model to more closely fit the full data sample at the expense of potentially larger bias. By the same token, the bias of $\bar x_{T+h}(\beta,\bar\theta(T,\beta))$ goes down as the randomization rate $\beta$ drops, thereby alleviating possible over-fitting of the full data sample at the expense of potentially larger variance. As in other ML methods, there is no impact on the irreducible variance term reflecting the inherent uncertainty about future states $x_{T+h}$ also when given full information up to time T.

Thus, the bias-variance decomposition in Proposition \ref{eq::MSE_decomposition} implies that for minimizing the expected prediction error under square loss function it should be beneficial to use the RMD framework whenever full utilization of all available data (i.e. $\beta=1$) entails a non-negligible bias term due to data over-fitting and model mis-specification in the presence of outliers. In such cases a data-driven estimate of the optimal value of $\beta \ll 1$ can be obtained using standard cross-validation techniques. By contrast, under the null of an unbiased model the zero bias term in Proposition \ref{eq::MSE_decomposition} would imply an optimal value of $\beta=1$.

From this perspective, in case of attaining an optimal $\beta < 1$ the RMD framework offers an easy approach to attaining model-free improvement when filtering and forecasting in the presenece of outliers based on any model of choice, e.g. preferred due to simplicity. In case of attaining an optimal $\beta \approx 1$ the RMD framework offers an easy approach to validating unbiasedness of the considered model. If one further takes the view that ``all models are wrong but some are useful'' then the RMD framework can thus offer an advantageous tool to help achieve model parsimony in applications of state-space models to forecasting (Occam's Razor).

To broaden the scope of these insights beyond the popular special case of a square loss function, it is useful to note that the bias-variance decomposition in Proposition \ref{eq::MSE_decomposition} readily extends also to a wide range of other loss functions. First, \citet{James_Hastie_1997} and \citet{James_2003} have shown how to obtain a straightforward generalization of the bias-variance decomposition in Proposition \ref{eq::MSE_decomposition} for any symmetric loss function. Second, \citet{Heskes_1998}, \citet{Wu_Vos_2012} and \citet{Wu_Vos_2015} have further obtained an analogous bias-variance decomposition valid for any kind of error measure that can be derived from a Kullback-Leibler divergence or loglikelihood stemming from the underlying probability model. Therefore, the above insights regarding the RMD-induced bias-variance tradeoff as a function of the randomization rate $\beta$ readily extend for any symmetric loss function as well as for the case when the expected prediction error captures the difference between the entire predictive distribution of $\bar x_{T+h}(\beta)$ and the respective target distribution of $x_{T+h}$ and associated likelihoods.

\subsection{Distinction from Related Work}

\subsubsection*{Relationship to Sim's Rational Inattention}

Similar to the idea of rational inattention by \citet{Sims_2003} and \citet{Sims_2011}, the RMD framework allows for the possibility that the optimal strategy for estimation and forecasting is to not fully fit to all observed data.
However, rather than supposing an exogenous cost to including more data, the RMD framework considers the risk of information loss from overfitting to misspecified values as an implicit cost of fitting to all observed data.
This gives rise to ``randomized innatention'' to measurements and an optimal randomized usage rate of the data to a degree where the marginal loss of including a possibly misspecified observation, e.g., an outlier, equals the marginal gain of including it.
While in practice the true expectation of information gain and loss cannot be calculated, this tradeoff can be achieved by setting the hyperparameter $\beta = P(C_t)$, via cross-validation or other related methods, to optimize  a chosen loss function such as the log-forecasting density or Mean Squared Forecasting Error (MSFE) at a set forecasting horizon.

\subsubsection*{Relationship to Heavy Tailed Modelling}

Assuming that the choice of $f_j$ is more diffuse than the observation kernel $f$, the implementation of RMD naturally induces a heavier-tailed measurement equation, which has precedence in the literature to make state space models more robust to outliers. Models with a Gaussian measurement are particularly susceptible to outliers, and often need to be adapted to achieve robustness in applications. Such adaptations include using a mixture of Gaussians as in \citet{Kitagawa_1989} or a $T$ distribution as in \citet{Harvey_Luati_2014} to better approximate the true heavy-tailed measurement.
However, an advantage of the RMD framework is that it can be applied to any state-space model to safeguard from misspecification, even in cases where there are no obvious heavy-tailed modifications to the observation kernel.
Moreover, existing robustification methods work by directly modifying $f$ to allow for more extreme realizations, and these methods are not as flexible as the RMD framework since they usually assume that the degree of kurtosis exhibited by $f$ is constant conditional on some hidden states and parameters.

\subsubsection*{Relationship to Machine Learning Methods}

The RMD-X implementation of the RMD framework effectively can be viewed as a time-series extension of bootstrap aggregation (bagging), originally developed by \citet{Breiman_1996}. As a key distinction from bagging, RMD-X preserves time-series dependence by re-sampling without replacement while retaining the original time index of each observation by randomly drawing only induced missing values in each re-sampled path.
This latter feature is a key difference from the extant literature on bagging, which focuses mostly on cross-sectional settings rather than time-series models where temporal dependence puts natural restrictions on resampling methods.
A handful of studies that have considered the benefits of bootstrapping and bagging in a time-series context have done so either by resampling the full-sample model-based i.i.d. residuals to assess estimation uncertainty such as \citet{Stoffer_Wall_1991} or by resampling entire blocks of dependent observations to robustify forecasts as in \citet{Inoue_Kilian_2008} and \citet{Bergmeir_Hyndman_Benitez_2016}.

\nocite{Fan_Ma_Wang_Zhu_2021}
\nocite{Fan_Li_Zhang_Zou_2020}

From a more general machine learning perspective, the RMD framework introduces a new bias-variance trade-off by varying data utilization through the randomization rate $\beta$. This relates to but is different from a wide range of machine learning methods exploiting such bias-variance trade-off. As a key distinction, the RMD framework aims to reduce prediction bias at the expense of larger variance by way of limiting data utilization, while many other ML methods aim to reduce prediction variance at the expense of larger bias by way of imposing penalization over the entire data sample as controlled by regularization parameters. Despite this inherent difference, the fraction of randomly utilized measurements in the RMD framework can still be viewed from a machine learning standpoint as a regularization parameter $\beta$ that controls the arising robustness-efficiency trade-off in the presence of misspecified observations with favorable learning properties. This further means that RMD filtering and forecasting could in principle be used also in combination with other penalty-based ML methods.

\section{Empirical Illustration: Inflation forecasting \label{sec:empirical}}

To illustrate the empirical performance of RMD filtering and forecasting we choose a well-known setting from the time-series literature on extracting inflation trends where the use of standard state-space models is known to suffer from the presence of measurement outliers and other sources of model mis-specification. This time-series setting offers an ideal controlled experiment allowing us to compare the empirical performance of RMD filtering and forecasting when applied to standard space models vis-a-vis the empirical performance of other modelling approaches that have gained popularity for robust extraction of inflation trends in the presence of outliers.

\subsection{Unobserved Components Model} \label{subsec:UC}

Market participants and central bankers alike traditionally pay considerable attention to data-driven assessments of long-run inflation expectations.
A large part of the literature on inflation forecasting has considered alternative econometric approaches to estimating inflation trends based on time series modelling of officially released price index data.\footnote{Another large strand of the literature aims to exploit the predictive content of collected survey-based measures of inflation. For a recent literature survey on inflation forecasting see for example \citet{Faust_Wright_2013}.}
\citet{Stock_Watson_2007} and \citet{Stock_Watson_2016} provide compelling evidence for time-variation in the precision of inflation rate measurements as well as the presence of additional measurement distortions due to outliers.
Inspired by these findings, we consider the ability of our robust filtering approach to successfully guard against the impact of inflation measurement imperfections without the need to explicitly model them for the purposes of improved forecasting of long-run inflation trends.
In particular, we apply our RMD framework to the classical unobserved components (UC) representation of the IMA(1,1) benchmark model for inflation forecasting scrutinized also by \citet{Stock_Watson_2007}. This RMD augmented specification is given by the following set of equations:

\begin{align}
  x_t | x_{t-1},\theta  &\sim g(. |x_{t-1}) = N(x_{t-1},\sigma_{\varepsilon}^2) \tag{\ref{eq::SSM_state}'}\label{eq::SSM_state_inflation} \\
  y_t |x_t,\theta &\sim f(. |x_t) = N(x_t,\sigma_{\eta}^2) \tag{\ref{eq::SSM_obs_prec}'}\label{eq::SSM_obs_prec_inflation} \\
  {y}_t^{(j)}      &\sim f_{j} (.) \tag{\ref{eq::SSM_junk_value}'}\label{eq::SSM_junk_value_inflation}  \\
  C_t    &\sim  Bernoulli(\beta) \tag{\ref{eq::SSM_junk_indicator}'}\label{eq::SSM_junk_indicator_inflation}  \\
           \hat {y}_t           &= \begin{cases} \tag{\ref{eq::SSM_missing_cond}'}\label{eq::SSM_missing_cond_inflation}
             y_t ~ \text{if} ~ C_t = 1 \\
             y_t^{(j)} ~ \text{if} ~ C_t = 0
           \end{cases}
\end{align}
where $\hat y_t$ denotes the observed change in the log price-level (PCE) at quarter $t$ subject to potential distortions tackled by the RMD equations (\ref{eq::SSM_junk_indicator_inflation})-(\ref{eq::SSM_missing_cond_inflation}). $C_t$ is the Bernoulli indicator variable which controls whether $\hat y_t$ is informative about the underlying rate $\beta \in [0, 1]$. In this RMD augmented setting, $y_t$ is the undistorted and hence informative measurement which depends on $x_t$, the latent trend component of inflation, and model parameters $\theta := (\sigma_\eta, \sigma_\epsilon)$. In contrast, $y_t^{(j)}$ is the alternative distorted measurement which contains no information about $x$ or the model parameters.
We consider both the RMD-N and RMD-X implementations of the RMD framework. In the case of RMD-N, as described in Section \ref{subsec:RMD_N_Assumption}, we set the alternative distribution $f_{j}$ as $F(\hat y_t | Y^{t-1}, \beta) := \int f(\hat y_t | x_t, \theta ) dP(x_t, \theta | Y^{t-1})$, the one-step-ahead predictive density. In the case of RMD-X, we consider the simplified implementation given in Section \ref{subsection:RMD_X}.

\begin{figure}[h]

  \noindent{\caption{Filtered mean comparison between $\beta=1$ (dashed red) and $\beta=0.15$ (solid purple) for RMD-X applied to the Unobserved Components (UC) model of Stock \& Watson (2007), where black dots represent the observed log-quarterly inflation. \label{fig:uc_filtered_rmdx_realdata}}}
  \begin{center}
    \begin{tabular}{c}
      \includegraphics[height=2.86in]{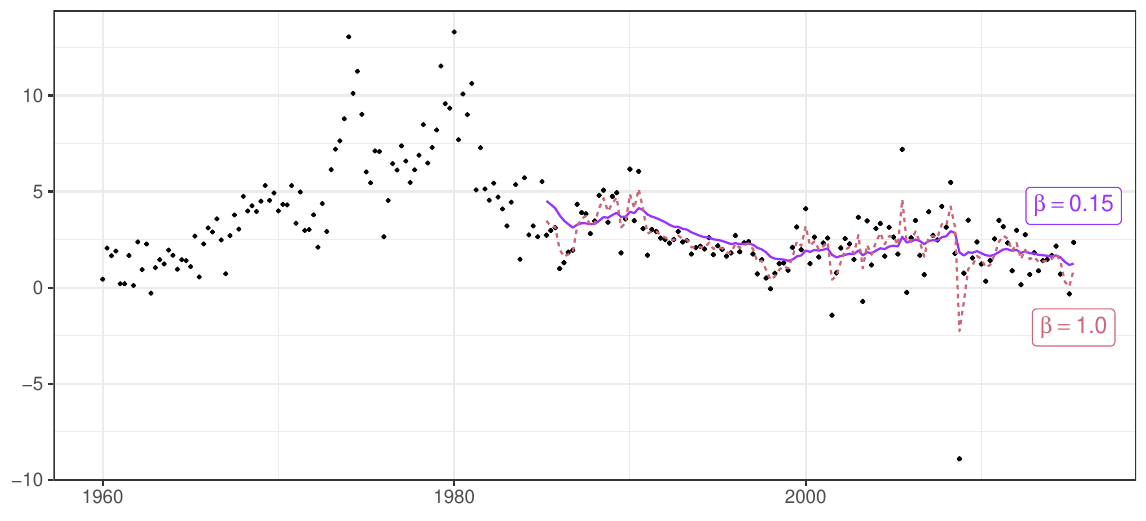}
    \end{tabular}
  \end{center}
\end{figure}

\begin{figure}[h]

  \noindent{\caption{Filtered mean comparison between $\beta=1$ (dashed red) and $\beta=0.15$ (solid blue) for RMD-N applied to the Unobserved Components (UC) model of Stock \& Watson (2007), where black dots are the observed log-quarterly inflation. The smoothed posterior probability $P(C_t = 1 | \hat y^T)$ for $\beta = 0.15$ is shown below. Large black dots indicate observations for which $P(C_t = 1| \hat y^T) < 0.005$.  \label{fig:uc_filtered_realdata}}}
  \begin{center}
    \begin{tabular}{c}
      \includegraphics[height=3.5in]{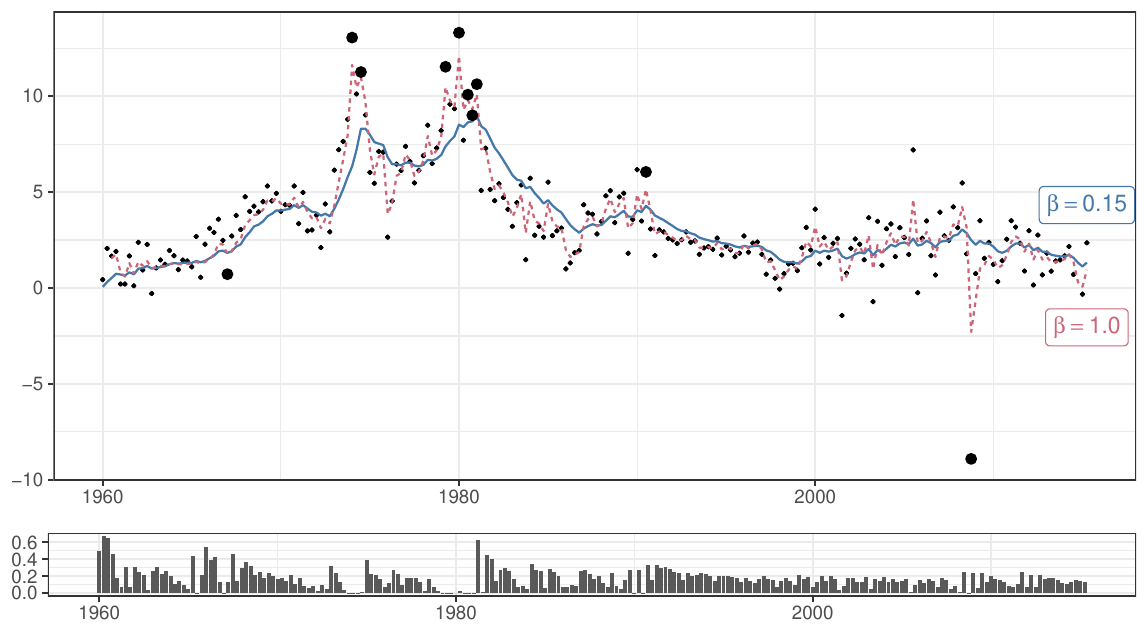}
    \end{tabular}
  \end{center}
\end{figure}

We estimate this model with the same aggregate PCE price index quarterly data series (PCE-all) used in \citet{Stock_Watson_2016} from 1960Q1 to 2015Q2 with a forecast evaluation period set to 1990Q1-2015Q2, in order to directly compare with the more recent UCSVO model presented in that paper.\footnote{We thank \citet{Stock_Watson_2016} for making publicly available the data and program codes necessary for replicating their results.} First, to illustrate how RMD impacts the estimation of latent states, we compare the filtered mean from the original UC model ($\beta = 1$) and the UC model augmented with either RMD-X (\autoref{fig:uc_filtered_rmdx_realdata}) or RMD-N (\autoref{fig:uc_filtered_realdata}) at ($\beta = 0.15$).
Without applying missing data randomization, the filtered mean for the original UC model visually overfits the observed inflation in each quarter.
On the other hand, the filtered mean for the UC models augmented with RMD-X and RMD-N appear to follow a smoother path which better tracks the long-run trend of the process.
This supports the notion in Section 3 that RMD favors smoother dynamics which follow long-term trend at the cost of short-term fit.
Focusing on RMD-N specifically, the bottom panel of \autoref{fig:uc_filtered_realdata} shows the inferred posterior smoothed probability that $C_t = 1$ at each point in time using all available data.
The model identifies several points (indicated as large black dots) which are very unlikely to have information about the unobserved component ($P(C_t < 0.005| \hat y^T)$), and these points coincide with what one might visually determine as outliers.
However, rather than relying on strict \textit{apriori} assumptions to filter outliers, these points are identified simply from the dynamics of the process.
Specfically, if future values do not follow the trend implied by an observation, the smoothed probability $P(C_t = 1 | \hat y^{T})$ decreases, allowing for an interpretable mechanism to filter our or downweight contaminated measurements.

\begin{table}[h]
\centering
\caption{UC model parameters estimated with RMD-N for different values of $\beta$ based on data from 1960Q1 to 2015Q2. \label{tbl:params_uc_realdata}}
\renewcommand*{\arraystretch}{0.75}
\begin{tabu}{llrrr}
  \toprule
  \vspace{1mm}
$\beta$ & Posterior Percentile & $\sigma_\epsilon$ & $\sigma_\eta$ \\
\midrule
0.15  & 50\%               &  0.350 & 0.144 & \\
\rowfont{\scriptsize}
& 2.5\%               &  0.191 & 0.000 & \\
\rowfont{\scriptsize}
& 97.5\%              &  0.641 & 0.781 & \\
\midrule
0.25  & 50\%               &  0.360 & 0.435 & \\
\rowfont{\scriptsize}
& 2.5\%               &  0.222 & 0.002 & \\
\rowfont{\scriptsize}
& 97.5\%              &  0.645 & 0.870 & \\
\midrule
0.90  & 50\%               &  0.627 & 1.068 & \\
\rowfont{\scriptsize}
& 2.5\%               &  0.445 & 0.868 & \\
\rowfont{\scriptsize}
& 97.5\%              &  0.841 & 1.290 & \\
\midrule
1.00  & 50\%               &  0.653 & 1.375 & \\
\rowfont{\scriptsize}
& 2.5\%               &  0.467 & 1.205 & \\
\rowfont{\scriptsize}
& 97.5\%              &  0.891 & 1.564 & \\
\bottomrule
\end{tabu}
\end{table}
\begin{figure}[h]
  \noindent{\caption{Posterior distributions of UC model parameters estimated with RMD-N for different values of $\beta$ based on quarterly inflation data from 1960Q1 to 2015Q2. From lightest to darkest, the shaded regions correspond to the middle 50\%, 90\% and 98\% of the distribution. \label{fig:uc_posterior_parameters_realdata}}}
  \scriptsize \vspace{1mm}
  \vspace{1mm}
  \begin{footnotesize}
    \begin{singlespace}
    \end{singlespace}
  \end{footnotesize}
  \vspace{1mm}
  \begin{center}
    \begin{tabular}{c}
      \includegraphics[height=3.5in]{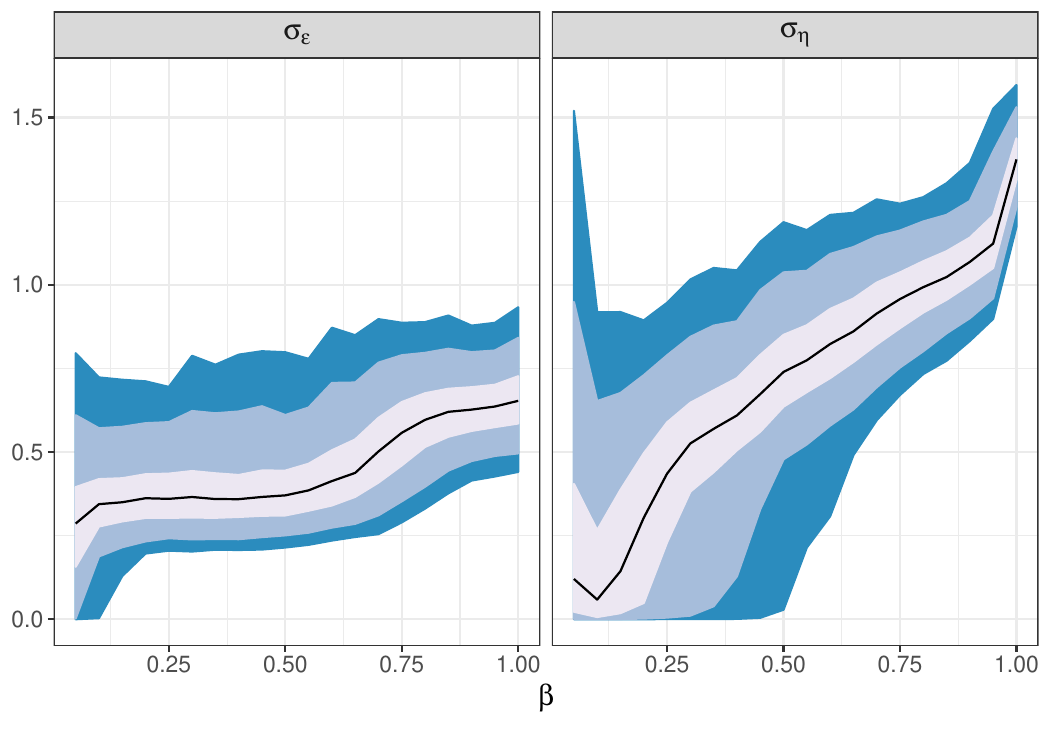}
    \end{tabular}
  \end{center}
\end{figure}

Next, to demonstrate the change in estimated parameters from applying RMD, Table \ref{tbl:params_uc_realdata} presents parameter estimates for the UC model with RMD-N using varying values of $\beta$ estimated over the full sample from 1960Q1 to 2015Q2.
It is evident that both the observation and state equation variance increases with the increase of $\beta$.
This is explained by the requirement of the UC model to fit increasingly more and larger outliers present in the data.
The increase is more pronounced for the parameter $\sigma_{\eta}$ that controls the precision of the observation equation.
To better illustrate the changes in the estimates of these parameters as we change the randomization parameter $\beta$ we also present the plots with their estimates in Figure \ref{fig:uc_posterior_parameters_realdata}.
As it is evident from Figure \ref{fig:uc_posterior_parameters_realdata}, the low and medium values of $\beta$ do not have a significant impact on the estimates of $\sigma_{\varepsilon}$, while the estimates of $\sigma_{\eta}$ grow steadily with the values of $\beta$.
Since higher values of $\beta$ naturally imply that more outliers are to be fitted by the model we should expect the variances of both the observation and state equations to increase with this parameter.
However, this effect is stronger for the observation equation.

\begin{table}[ht]
\centering
\caption{\textbf{Comparison of forecasting performance of the UC model with RMD-N for different values of $\beta$}. The UC model is estimated based on data from 1960Q1 to 2005Q2, evaluating forecast density at 1-,4-,8-, and 12-quarter-ahead horizons starting in 1990Q1. The Weighted Likelihood Ratio (WLR) test from \citet{Amisano_Giacomini_2007} reported in the table compares forecasts based on the UC model with RMD-N for different values of $\beta$ versus the ones for $\beta = 1$ standing for the original UC model as a benchmark. \label{tbl:WLR_UC_realdata}}
\begin{tabu}{llrrrr}
  \toprule
    \vspace{1mm}
 $\beta$ & $h$ & $\widehat{\text{WLR}}$ & $\hat{\sigma}_{\text{\tiny{WLR}}}$ & $t$ & $P$ \\
\midrule
0.15 & 1  & 0.2981 & 1.75 & 1.71 & 0.95 \\
    & 4   & 0.2661 & 0.86 & 3.08 & 0.99 \\
    & 8   & 0.2593 & 0.45 & 5.56 & 0.99 \\
    & 12  & 0.2688 & 0.33 & 7.85 & 0.99 \\
\midrule
0.25 & 1  & 0.2748 & 1.52 & 1.82 & 0.96 \\
    & 4   & 0.2507 & 0.73 & 3.39 & 0.99 \\
    & 8   & 0.2514 & 0.47 & 5.15 & 0.99 \\
    & 12  & 0.2688 & 0.37 & 6.93 & 0.99 \\
\midrule
0.90 & 1  & 0.0403 & 0.26 & 1.58 & 0.94 \\
    & 4   & 0.0222 & 0.16 & 1.36 & 0.91 \\
    & 8   & 0.0130 & 0.12 & 1.03 & 0.84 \\
    & 12  & 0.0170 & 0.19 & 0.84 & 0.80 \\
\bottomrule
\end{tabu}
\end{table}

To evaluate the forecasting performance of the UC model with RMD, we first consider whether the improvement over the original UC model is significant.
We use the \citet{Amisano_Giacomini_2007} test statistics, $\widehat{WLR}$, to compare the forecasting performance of the model augmented with RMD-N versus the original model.
The $\widehat{WLR}$ statistics are computed relative to the model with $\beta=1$ utilizing all inflation measurements with no randomization, which corresponds to the original UC specification in \citet{Stock_Watson_2007}.
Table \ref{tbl:WLR_UC_realdata} presents results for the UC model estimated recursively using the sample from 1960Q1 to 2005Q2 to generate forecast densities evaluated at 1-, 4-, 8-, and 12-quarter-ahead horizons starting from 1990Q1.
For brevity, we document the results for three different choices of the parameter $\beta$ of 0.15, 0.25 and 0.9.
It is evident that all specifications with data randomization perform better than the benchmark UC model producing positive t-statistics with statistically significant improvements for all horizons for the specifications with lower values of $\beta$ of 0.15 and 0.25.

After showing significant improvement in density forecasts, we compare the Mean Squared Forecasting Error (MSFE) of the UC model augmented with either RMD-X or RMD-N to that of the original UC model ($\beta = 1$). Following \citet{Stock_Watson_2007} and \citet{Stock_Watson_2016}, we target the average inflation over 1-, 4-, 8-, and 12-quarters ahead.
For completeness, we apply these criteria also to two other existing approaches for attaining robustness to outliers as alternative benchmarks against which we evaluate the relative performance of RMD.
The first approach, which we denote as UC-T, follows \citet{Harvey_Luati_2014} in replacing the Gaussian measurement in \autoref{eq::SSM_obs_prec_inflation} with a scaled $T$ distribution with $\nu > 0$ degrees of freedom.
The second approach is the unobserved components/stochastic volatility outlier-adjustment (UCSVO) model representing a more recent extenstion of the UC model by \citet{Stock_Watson_2016}. The UCSVO model also minimizes the detrimental impact of outliers by subjecting them to particular distributional assumptions while also allowing for stochastic volatility.
\begin{figure}[h]
  \noindent{\caption{Optimal value of $\beta$ selected over time using information up to a given date. Each box compares $\beta$ selected by RMD-X and RMD-N optimizing a given forecasting MSFE criterion (1, 4, 8, or 12-quarters ahead) for different underlying models (UC, AR, or ARMF). \label{fig:all_beta_realdata}}}
  \begin{center}
    \begin{tabular}{c}

      \includegraphics[height=5in]{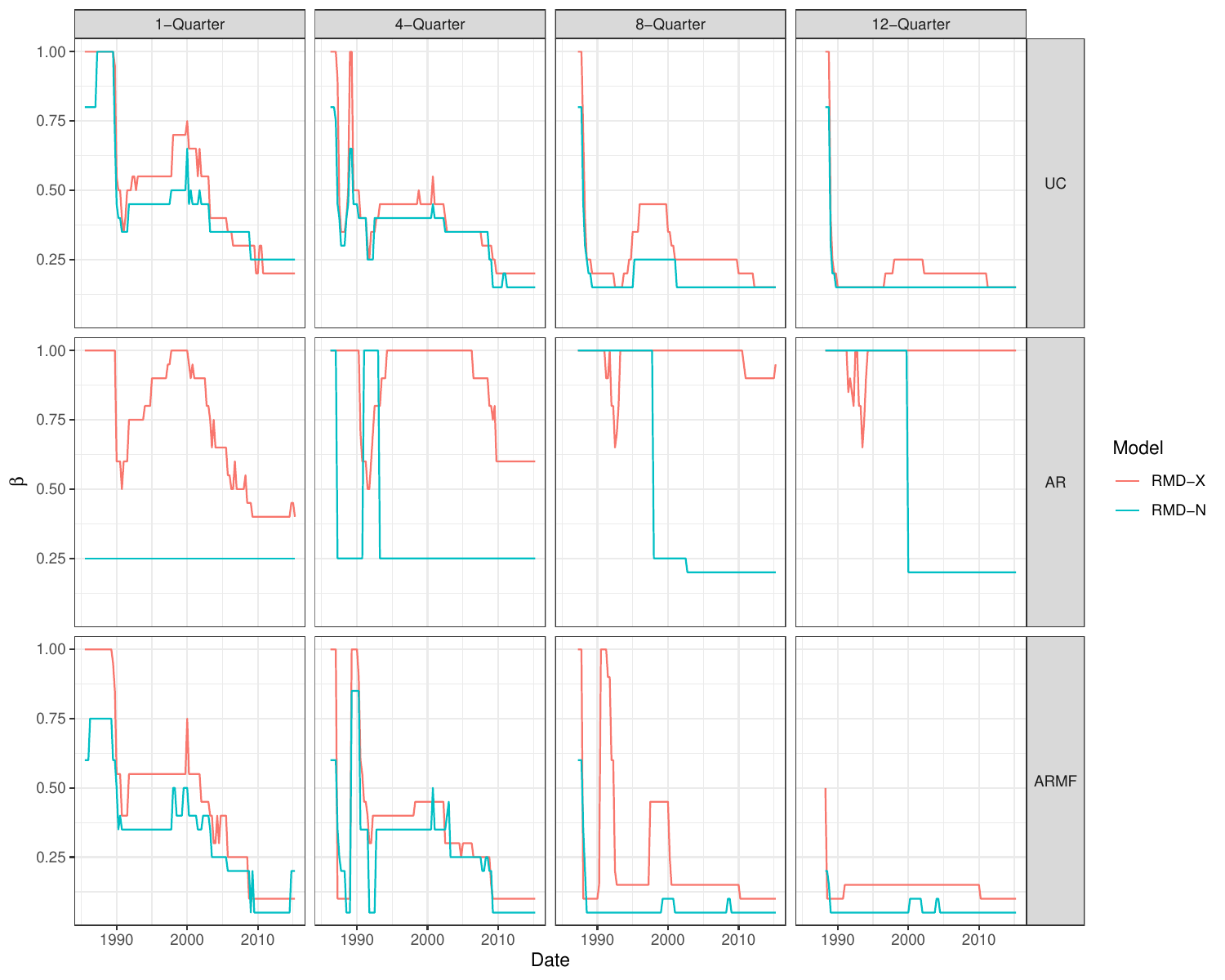}
    \end{tabular}
  \end{center}
\end{figure}

\begin{table}
\centering
\caption{MSFE comparison for different considered state space models for extracting inflation trends augmented with RMD-X and RMD-N for a range of values of $\beta$ optimized recursively to different forecasting horizons, compared to the UC and UCSVO benchmark models from Stock \& Watson (2007) and Stock \& Watson (2016) respectively. The considered forecast horizons are: one quarter (Q1), four quarter (Q2), eight quarter (Q8) and twelve quarters (Q12).}
\label{tbl:msfe_perf_all}
\begin{tabular}{llr|rrrr}
  \toprule
 Model & RMD Approach & Strategy for $\beta$ & Q1 & Q4 & Q8 & Q12 \\
  \midrule
    UC & RMD-X &   Q1         & 2.96 & 1.14 & 0.86 & 0.82 \\
       &      &   Q4         & 2.95 & 1.10 & 0.84 & 0.80 \\
       &      &   Q8         & 2.94 & 1.02 & 0.74 & 0.66 \\
       &      &  Q12         & 3.03 & 1.04 & 0.74 & 0.68 \\
       & N/A  & $\beta = 1$ & 3.33 & 1.58 & 1.26 & 1.22\\
  \midrule
       & RMD-N &   Q1         & 2.92 & 1.13 & 0.85 & 0.78 \\
       &      &   Q4         & 2.93 & 1.11 & 0.85 & 0.78 \\
       &      &   Q8         & 2.88 & 0.97 & 0.67 & 0.59 \\
       &      &  Q12         & 2.96 & 0.98 & 0.66 & 0.59 \\
       & N/A & $\beta = 1$ & 3.33 & 1.58 & 1.26 & 1.22\\
  \bottomrule
  \toprule
    AR & RMD-X &   Q1         & 3.16 & 1.46 & 1.44 & 1.75 \\
       &      &   Q4         & 3.34 & 1.54 & 1.49 & 1.79 \\
       &      &   Q8         & 3.44 & 1.61 & 1.53 & 1.84 \\
       &      &  Q12         & 3.55 & 1.66 & 1.59 & 1.92 \\
       & N/A  & $\beta = 1$ & 3.31 & 1.61 & 1.48 & 1.74\\
  \midrule
       & RMD-N &   Q1         & 3.38 & 1.80 & 2.13 & 3.09 \\
       &      &   Q4         & 3.50 & 1.79 & 2.10 & 2.98 \\
       &      &   Q8         & 3.97 & 2.19 & 2.17 & 2.44 \\
       &      &  Q12         & 4.36 & 2.40 & 2.19 & 2.24 \\
       & N/A  & $\beta = 1$ & 5.42 & 3.51 & 3.11 & 3.02\\
  \bottomrule
  \toprule
  ARMF & RMD-X &   Q1         & 2.89 & 1.01 & 0.64 & 0.52 \\
       &      &   Q4         & 3.01 & 1.11 & 0.77 & 0.60 \\
       &      &   Q8         & 2.96 & 1.00 & 0.62 & 0.44 \\
       &      &  Q12         & 2.98 & 0.92 & 0.51 & 0.33 \\
       & N/A  & $\beta = 1$ & 3.27 & 1.39 & 0.92 & 0.75 \\
  \midrule
       & RMD-N &   Q1         & 2.86 & 1.00 & 0.67 & 0.55 \\
       &      &   Q4         & 2.96 & 1.06 & 0.73 & 0.57 \\
       &      &   Q8         & 2.92 & 0.90 & 0.54 & 0.41 \\
       &      &  Q12         & 2.91 & 0.86 & 0.48 & 0.33 \\
       & N/A  & $\beta = 1$ & 3.23 & 1.31 & 0.84 & 0.66 \\
  \bottomrule
  \toprule
  \multicolumn{3}{c|}{Benchmarks} & Q1 & Q4 & Q8 & Q12 \\
  \midrule
  \multicolumn{3}{r|}{UC-T ($\beta = 1$)}        & 3.24 & 1.41 & 1.13 & 1.01 \\
  \multicolumn{3}{r|}{UCSVO ($\beta = 1$)}                 & --   & 1.09 & 0.81 & 0.69 \\
  \multicolumn{3}{r|}{Naive Forecast of 2\%} & 8.97 & 7.40 & 6.96 & 6.67 \\
\end{tabular}
\end{table}

\begin{figure}[H]
  \noindent{\caption{Mean Squared Forecasting Error (MSFE) over different horizons (lower values are better). Each box compares the performance of RMD-X (red solid lines) and RMD-N (green solid lines) using a given criterion for selecting $\beta$ (1, 4, 8, or 12-quarters ahead) and a given model (UC, AR, or ARMF). Latent states, parameters, and $\beta$ are estimated from 1960Q1 to 2015Q2, and the MSFE is calculated with forecasts formulated in 1990Q1 onward. MSFEs from the original model with $\beta = 1$ (dashed lines) and the UCSVO model (blue dots) from \citet{Stock_Watson_2016} are plotted for reference. \label{fig:all_msfe_strat_realdata}}}
  \begin{center}
    \begin{tabular}{c}
      \includegraphics[height=5in]{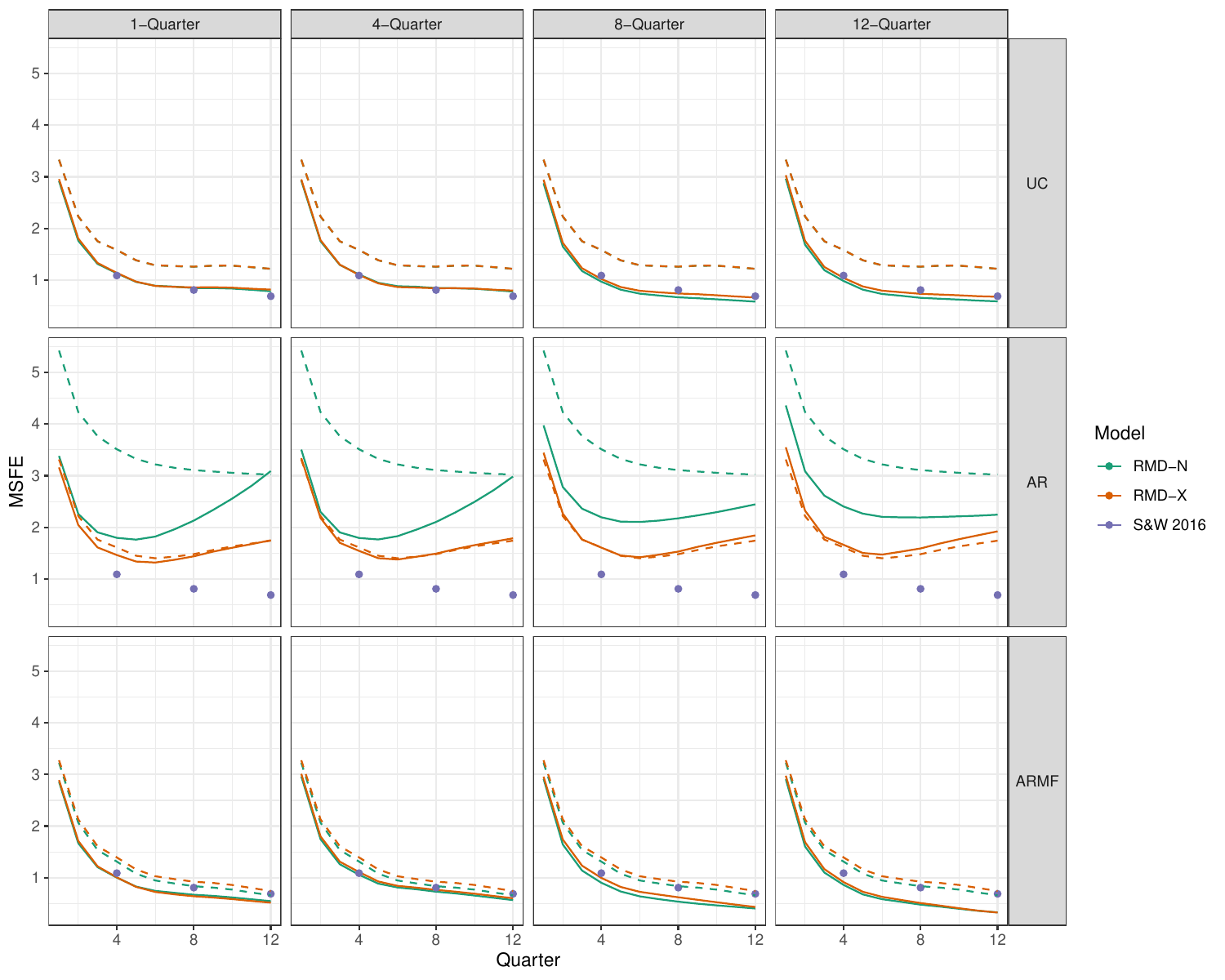}
    \end{tabular}
  \end{center}
\end{figure}

For a fair comparison, the latent states and parameters are re-estimated online at each point in time from 1960Q1 to 2015Q1 using only past information (no look ahead), with Mean Squared Forecasting Error (MSFE) evaluated across 1, 4, 8, and 12-quarter horizons from 1990Q1 onward.
Moreover, the hyperparameter $\beta$ is chosen at each point in time as the value minimizing MSFE in targeting average inflation over 1-, 4-, 8-, and 12-quarter horizons also based on past data up to that point in time with no look-ahead (see Figure \ref{fig:all_beta_realdata}, top row).
Thus, at each point in time and for each of the different optimizing criteria, the chosen optimal $\beta$ is used to formulate forecasts.
The resulting MSFE from each of these four strategies can then be used to objectively compare RMD-X and RMD-N to the UC-T and UCSVO benchmarks.
As shown in \autoref{tbl:msfe_perf_all} and in the top row of Figure \ref{fig:all_msfe_strat_realdata}, there is a clear improvement in adopting RMD-X or RMD-N over the base UC model, particularly when aiming to optimize over a 12-quarter horizon.
Specifically, RMD-N achieves a 51\% reduction in MSFE over the base UC model at 12-quarter horizons, as well as a 41\% reduction compared to the UC-T model.
Moreover, compared to UCSVO across the same 12-quarter horizon, the UC model augmented with RMD-N leads to a 14\% reduction in MSFE. These gains demonstrate the ability of our RMD approach to robustly extract the persistent inflation trend component in the UC model from noisy and potentially misspecified or outlier-contaminated inflation measurements without the need to parametrize such more complex features as done in the UCSVO model.
Overall, these results indicate that employing our RMD filtering framework significantly improves the forecasting performance of the classical UC model by naturally safeguarding against overfitting data outliers, thereby offering an attractive alternative to the UCSVO model.

\subsection{Generalized AR Model Reflecting Inflation Targeting} \label{sec:AR}

The transition equation of the state process in the UC model, as shown in equation (\ref{eq::SSM_state_inflation}), can be viewed as a corner case of the following more general autoregressive (AR) model:

\begin{equation} \label{eq::AR_state_inflation}
  x_t | x_{t-1},\theta  \sim g(x_t|x_{t-1}) = N(\mu + \kappa (x_{t-1} - \mu),\sigma_{\varepsilon}^2) \\
\end{equation}

Here, $\kappa$ reflects the degree of mean-reversion of the process, and the original UC model is equivalent to fixing $\kappa = 0$, which simultaneously removes $\mu$ as an identifiable parameter.
Inflation targeting long adopted by central banks further provides justification from economic theory for modelling inflation with a long-run or stationary mean.
For example, in the case of the United States an inflation target of 2\%
has been maintained by the Federal Reserve for many years (see \citet{FedInflation}).
However, even if the model were correctly specified, the long-run mean inflation rate
cannot be estimated precisely without a large sample if $\kappa \approx 0$ (i.e., the process is close to a random walk).
We therefore consider embedding knowledge about the inflation target policy in place directly into the model by fixing the long run rate at $2 \%$.
Thus, in addition to examining the performance of an unrestricted AR model, we also assess the forecasting performance of an alternative version of the same model (denoted ARMF) where $\mu$ is fixed a priori at the $2 \%$ inflation target.

When applying our RMD framework to the AR model as we did for UC, \autoref{fig:all_beta_realdata} shows that RMD-X selects $\beta = 1$ for most $t$ at longer horizons for the unrestricted AR model with hard to estimate long-run mean. The RMD-N model favors a lower $\beta$ and shows improvement over the base model, but the forecasting performance of RMD-N is worse than that of RMD-X across the board.
This discrepancy is likely due to the fact that, as a fully Bayesian estimation, RMD-N includes the large uncertainty about $\mu$ in its forecasts, which is detrimental to forecasting performance.
Neither implementation of RMD reaches the performance of the benchmark UCSVO model. These findings are consistent with results from the prior literature also favoring UC over AR models for extracting inflation trends.

On the other hand, explicitly reflecting the inflation target in place by way of imposing a long-run mean of $2\%$ in the ARMF model vastly improves forecasting performance compared to the UC model, even without using the RMD framework. Furthermore, applying the RMD-N framework to ARMF leads to a substantial improvement in the MSFE performance in the 12-quarter horizon over both the UCSVO benchmark (52\% reduction) and that of the UC model augmented with RMD-N (43\% reduction). These improvements in inflation forecasting can also be taken as new evidence that inflation targeting has been effective. \footnote{As a sanity check, we also compared forecasting results to a trival plug-in method where we set the $2\%$ inflation target for every forecast. The performance of this naive method was significantly worse than that of any of the considered state space models. See last row in \autoref{tbl:msfe_perf_all}.}

Thus, in all three considered state space models for inflation (UC, AR and ARMF), adopting our RMD filtering framework  consistently leads to often notable degrees of improvement in forecasting performance.
Moreover, based on the vast RMD improvements attained for the UC and ARMF models which impose specific parameter constraints, we note that the application of the RMD framework can be beneficial also after imposing additional sources of information from economic theory.
As RMD has the effect to regularize the estimated latent states toward path dynamics dictated by $\theta$, it should not be surprising that having extra information about $\theta$ \textit{a priori}, e.g. stemming from economic theory, can be crucial for filtering and forecasting performance. This further underscores that expert modelling still remains important for successful application of machine learning techniques by leveraging also particular theory-implied features of the data generating process.

\section{Conclusions}\label{sec:Conclusions}

Similar to the justification for rational inattention (\citet{Sims_2003}, \citet{Sims_2011}), in this paper we argue that the presence of measurement outliers and misspecification in state-space models may limit the scope for utilizing or putting equal weight on every available data measurement. More specifically, we note that even absent any explicit costs to data collection and processing, the optimal degree of utilization of imperfect data would implicitly depend on loss function sensitivities to bias versus efficiency. This motivates us to put forward a randomized missing data (RMD) approach to robust filtering of state-space models aiming to induce a degree of ``randomized innatention'' to available measurements. We accomplish this by purposely randomizing over the utilized subset of seemingly highly precise but possibly misspecified or outlier contaminated data measurements in their original time series order, while treating the rest as if missing. Such randomization ensures that time-series dependence is fully preserved and all available measurements can get utilized subject to a degree of downweighting depending on the loss function of interest. It also exploits the idea that the inclusion of only a small fraction of available highly precise measurements can still extract most of the attainable efficiency gains for filtering latent states, estimating model parameters, and producing out-of-sample forecasts. The arising robustness-efficiency trade-off is controlled by varying the fraction of randomly utilized measurements or the incurred relative efficiency loss from such randomized utilization of the available measurements.

On the theory side, we first lay down the general properties of our RMD framework in a Bayesian setting and then provide two alternative implementations. The first one, denoted RMD-N, allows for endogenous randomization of missing data based on an indicator that is subject to a learning assumption. The second implementation, denoted RMD-X, is subject to a no-learning assumption with exogenous randomization of missing data. We establish that both RMD-N and RMD-X arise as special cases of our general RMD framework under specific convenient choices of the underlying conditional distribution of the missing or misspecified observations. Our main theory result in Section \ref{sec::Bias_Variance_Decomposition} above is that the RMD framework introduces a new bias-variance trade-off controlled by the induced randomization rate. As a key distinction, the RMD framework limits data utilization to reduce prediction bias at the expense of larger variance, while a wide range of ML methods exploit bias-variance trade-off involving penalization over the entire data sample to reduce prediction variance at the expense of larger bias.
The RMD-X implementation of the RMD framework can further be viewed as a time-series extension of bootstrap aggregation (bagging), originally developed by \citet{Breiman_1996}. As a key distinction from bagging, RMD-X preserves time-series dependence by re-sampling without replacement while retaining the original time index of each observation by randomly drawing only induced missing values in each re-sampled path.

On the empirical side, we show consistently attractive performance of our RMD framework and the resulting out-of-sample forecasts in popular state space models for extracting inflation trends along with model extensions that more directly reflect inflation targeting by central banks. First, we demonstrate that RMD-N and RMD-X augmented versions of the classical UC model offer a viable alternative to the well-established unobserved components/stochastic volatility outlier-adjustment (UCSVO) model that was put forward by \citet{Stock_Watson_2016} and can improve forecasting average inflation, especially over longer horizons such as twelve quarters ahead. Second, we document even greater forecast improvements over a UC model with t-distributed innovations without stochastic volatility (UC-T), as another indication that the filtering and forecasting benefits offered by RMD cannot be attained by simply imposing a heavier-tailed measurement distribution.

Finally, we also consider a more general AR model for the unobserved inflation component which allows for reversion to a long-run mean either estimated from the data or fixed at the inflation target rate of $2\%$ as a way to more directly reflect central bank inflation targeting that has been in place for many years. The much superior forecasting performance we document when the mean is fixed to the \textit{a priori} $2\%$ inflation target in place  suggests that inflation targeting has been effective. The application of the RMD framework can be beneficial also after imposing additional economic information about the underlying model. This further underscores that expert modelling can still be important for successful application of machine learning techniques by leveraging also more specific theory-implied features of the data generating process. That said, both when applied to the classical UC and better-performing AR models with fixed inflation target mean our RMD approach to robust filtering and forecasting offers attractive empirical performance gains and ease of implementation in comparison to existing alternatives.

Overall, the proposed RMD framework for robust filtering and forecasting shows promising avenues for further exploration on both the theory and empirical side in a host of other areas of applications of state-space modelling and forecasting in the presence of data imperfections.

\newpage

\subsection*{Acknowledgements}

The authors thank Jeroen Dalderop, Ian Dew-Becker, Leland Farmer, Marco Del Negro, Bjorn Eraker, Jordi Llorens, Andrew Patton, Nicholas Polson, Neil Shephard, Tatevik Sekhposyan, and Jonathan Wright for very helpful discussions and comments. The authors also thank conference participants at the 2nd Workshop on Financial Econometrics and Empirical Modeling of Financial Markets, Kiel Institute for the World Economy, May 3-4, 2018, the 2018 NBER-NSF Seminar on Bayesian Inference in Econometrics and Statistics (SBIES), Stanford University, May 25-26, 2018, the 2018 NBER-NSF Time Series Conference, September 7-8, 2018, UCSD, the 2020 Midwest Finance Association (MFA) Annual Meeting, August 6-8, 2020, the 2021 Federal Forecasters Conference, May 6, 2021, the 13th Annual Society for Financial Econometrics (SoFiE) Conference, June 15-17, 2021, the Econometric Research in Finance (ERFIN) Workshop, September 17, 2021, the 2021 Meeting of the Federal Reserve System Committee on Econometrics, September 29, 2021, the 15th International Conference on Computational and
Financial Econometrics (CFE), December 18-20, 2021, the Vienna-Copenhagen Conference on Financial Econometrics, June 2-4, 2022,
as well as seminar participants at the Federal Reserve Board of Governors, Federal Reserve Bank of Boston, Northwestern University and the U.S. Census Bureau.

\newpage

\appendix

\begin{singlespace}
\clearpage
\bibliographystyle{apalike}

\end{singlespace}
\newpage

\begin{appendices}
\numberwithin{equation}{section}

\section{RMD: Posterior distribution}\label{sec:RMD_posterior}

We start with the most general form for $y^{(j)}_t \sim f_j^t$, where $y^{(j)}_t$ is independent of $\theta$ and $x^T$ conditional on all information up to time $t-1$ (i.e. $\hat y^{t-1}$).
Within a Bayesian paradigm, independence from $\theta$ is well-defined
through the choice of prior.
The goal is to sequentially infer the joint posterior distribution $P(\theta, x^T, C^T | \hat y^{T}, \beta)$ as observations are added.
Specifically, we show how one can update the posterior $P(x^{t-1}, \theta | \hat y^{t-1}, \beta)$ after observing $\hat y_t$ to obtain $P(x^t, \theta | \hat y^t, \beta)$. From here on, we assume that the unconditional probability $P(C_t = 1)$ is $\beta$ and that $C_t$ is independent of all past random variables and $x_t, \theta$. Thus, $P(C_t = 1 | \hat y^{t-1}, x^t, \theta, \beta) = \beta$. We then derive the posterior distribution of $P(C_t = 1 | \hat y^t, \beta)$ using Bayes' rule:

\begin{equation} \label{eq::post_ct1}
  P(C_t = 1 | \hat y^t, \beta) = \frac{\beta P(\hat y_t | C_t = 1, \hat y^{t-1})}{\beta P(\hat y_t | C_t = 1, \hat y^{t-1}\beta) + (1 - \beta) P(\hat y_t | \hat y^{t-1},  \beta)}
  \end{equation}

From the setup, the distribution of $\hat y_t$ conditional on $(C_t, x^t, \theta, \hat y^{t-1}, \beta)$ is either $f_\theta$ if $C_t = 1$ or $f_j^t$ if $C_t = 0$. Define $F(y_t | \hat y^{t-1}, \beta) := \int_{x_t, \theta} f_\theta (y_t | x_t) dP(x_t, \theta | \hat y^{t-1}, \beta)$, which is the one-step-ahead predictive distribution of $y_t$ given $\hat y^{t-1}$ and $\beta$. \autoref{eq::post_ct1} can be rewritten directly using $F$ and $f_j$.

\begin{equation} \tag{\ref*{eq::post_ct1}'} \label{eq::post_ct2}
  P(C_t = 1 | \hat y^t, \beta) = \frac{\beta F(\hat y_t | \hat y^{t-1}, \beta)}{\beta F(\hat y_t | \hat y^{t-1}, \beta) + (1 - \beta) f_j(\hat y_t | \hat y^{t-1}, \beta)} =: \hat \beta(\hat y^t)
\end{equation}

We then focus on the conditional posterior distribution $P(x^t, \theta | C_t = i, \hat y^t, x^t, \theta, \beta)$ for $i \in \{0,1\}$. To do so, we consider the joint distribution with the true measurement $y_t$ included as well.

\begin{equation} \label{eq::post_dist_cond_ct}
  \begin{split}
    P(x^t, \theta | C_t, \hat y^t, \beta) &= \int_{y_t} P(x^t, \theta | y_t, C_t, \hat y^t, \beta) dP(y_t | C_t, \hat y^t, \beta) \\
    &= \int_{y_t} P(x^t, \theta | y_t, C_t, \hat y^{t-1}, \beta) dP(y_t | C_t, \hat y^t, \beta) \\
    &= \Big ( \int_{y_t} \frac{f_\theta(y_t | x_t, \theta)}{F(y_t | \hat y^{t-1}, \beta)}dP(y_t | C_t, \hat y^t, \beta) \Big ) P(x^t, \theta | C_t, \hat y^{t-1}, \beta)  \\
    &= \Big ( \int_{y_t} \frac{f_\theta(y_t | x_t, \theta)}{F(y_t | \hat y^{t-1}, \beta)}dP(y_t | C_t, \hat y^t, \beta) \Big ) P(x^t, \theta | \hat y^{t-1}, \beta)  \\
\end{split}
\end{equation}

The value of the integral depends on the value of $C_t$. If $C_t = 1$, then $y_t = \hat y_t$ almost surely, so $P(y_t | C_t = 1, \hat y^t)$ equals the Dirac probability measure at $\hat y_t$.
Consequently, the integral will evaluate to $\frac {f_\theta(\hat y_t | x_t, \theta)} {F(\hat y_t | \hat y^{t-1}, \beta)}$.
On the other hand, if $C_t = 0$, then, because $\hat y_t = y_t^{(j)}$ almost surely and $y_t^{(j)}$ is conditionally independent of $y_t$, $P(y_t | C_t = 0, \hat y_t, \beta) = P(y_t | \hat y^{t-1}, \beta) = F(y_t | \hat y^{t-1}, \beta)$.
Since this cancels out with the denominator, and the numerator is a probability density, the integral will evaluate to $1$. We then substitute these values into \autoref{eq::post_dist_cond_ct}:

\begin{equation} \tag{\ref*{eq::post_dist_cond_ct}'} \label{eq::post_dist_cond_ct_simplified}
  P(x^t, \theta | C_t, \hat y^t, \beta) = \begin{cases}
    \frac{f_\theta (\hat y_t | x_t, \theta)}{F(\hat y_t| \hat y^{t-1}, \beta)} P(x^t, \theta | \hat y^{t-1}, \beta)
    ~ \text{if} ~ C_t = 1 \\
    P(x^t, \theta | \hat y^{t-1}, \beta) ~ \text{if} ~ C_t = 0
  \end{cases}
\end{equation}

These findings confirm the intuition described earlier.
If $C_t = 1$, then the observed $\hat y_t$ is treated just like it were the true measurement $y_t$.
If $C_t = 0$, then $\hat y_t$ is discarded and the posterior belief about $x^t$ and $\theta$ remains unchanged.
Combining Equations \ref{eq::post_ct2} and \ref{eq::post_dist_cond_ct_simplified}, the full update of the posterior distribution of $x^t, \theta$ with the information $\hat y_t$ is characterized below:

\begin{equation} \label{eq::posterior_sequential}
  P(x^t, \theta | \hat y^t, \beta) = \Big (\hat \beta(\hat y^t) \frac{f_\theta(\hat y_t | x_t)} {F(\hat y_t | \hat y^{t-1}, \beta)} + (1 - \hat \beta(\hat y^t)) \Big ) g_\theta(x_t | x_{t-1})P(x^{t-1}, \theta | \hat y^{t-1}, \beta)
\end{equation}

Recursing back to $t=0$, the posterior distribution can be written as:

\begin{equation} \label{eq::posterior_full}
  P(x^T, \theta | \hat y^T, \beta) = \prod_{t=1}^T \Big (\hat \beta(\hat y^t) \frac{f_\theta(\hat y_t | x_t)} {F(\hat y_t | \hat y^{t-1}, \beta)} + (1 - \hat \beta(\hat y^t) \Big ) g_\theta(x_t | x_{t-1}) \mu_\theta (x_0) P(\theta)
\end{equation}

We notice that the posterior distribution as characterized in Equations (\ref{eq::posterior_sequential}) and (\ref{eq::posterior_full}) resembles how the posterior distribution of the originally specified model would update with new information, and are in fact equal if $\beta = 1$. For $\beta \in (0, 1)$, we see that the resulting posterior distribution at each stage is exactly a mixture between including or excluding the observed $\hat y_t$, with weights governed by the posterior probability $\hat \beta$. This shows that the RMD framework introduces a natural and interpretable way to incorporate misspecfied measurements into a model by way of downweighting via randomization, with the degree to which a data point is added to the model given exactly by the belief that it corresponds to the correctly specified measurement equation.
\end{appendices}

\newpage
\end{sloppypar}
\end{document}